\documentclass{autart}   

\usepackage{graphicx}       
\usepackage{amsmath}
\usepackage{amsfonts}
\usepackage{enumerate}   
\usepackage{bm}   
\usepackage{amssymb}
\usepackage{chngcntr}
\usepackage{ dsfont }
\usepackage{apptools}
\usepackage[dvipsnames]{xcolor}
\usepackage{savesym}
\savesymbol{AND}
\usepackage{algorithmic}
\usepackage{algorithm}
\usepackage{url}

\newcommand{\hlc}[1]{#1}

\newcommand{\nd}{\text{\normalfont d}}
\newcommand{\cov}{\textsf{\footnotesize cov}}

\newcommand{\mf}{\mathbf}
\newcommand{\tr}{\textsf{\footnotesize tr}}

\newcommand{\diag}{\textsf{\footnotesize diag}}
\newcommand{\interior}{\textsf{\small int}}
\newcommand{\andset}{,\ }
\newcommand{\cset}{\mathbb{S}}
\newcommand{\mfb}[1]{\bar{\mf{#1}}}
\newcommand{\ses}{\Gamma}
\newcommand{\seq}{\gamma}
\newcommand{\seqq}{\nu}
\newcommand{\len}{\textsf{\small len}}
\newcommand{\sptwo}{\textsf{\footnotesize SP}^2}
\newcommand{\sstar}{\text{\fontsize{10}{0}$\star$}}

\def\RevOne#1{\textcolor{black}{#1}}
\def\RevThree#1{\textcolor{black}{#1}}
\def\RevSeven#1{\textcolor{black}{#1}}
\def\RevTen#1{\textcolor{black}{#1}}

\def\RevEleven#1{\textcolor{black}{#1}}

\begin{document}

\begin{frontmatter}

\title{Latency vs precision: \\stability preserving perception scheduling} 

\thanks[footnoteinfo]{This work was supported via projects PID2021-124137OBI00 and TED2021-130224B-I00
funded by MCIN/AEI/10.13039/501100011033, by ERDF A way of making Europe and by the
European Union NextGenerationEU/PRTR, by the Gobierno de Aragón under Project DGA T45-20R, by the Universidad de Zaragoza and Banco Santander, by the Consejo Nacional de Ciencia y Tecnología (CONACYT-Mexico) with grant number 739841.}
\thanks[footnoteinfo2]{\textcolor{red}{
This is the accepted version of the manuscript: R. Aldana-López, R. Aragüés, and C. Sagüés, ``Latency vs precision: Stability preserving perception scheduling," Automatica, vol. 155, p. 111123, 2023, ISSN 0005-1098. doi: 10.1016/j.automatica.2023.111123.
    \textbf{Please cite the publisher's version}. For the publisher's version and full citation details see:\\\protect\url{https://doi.org/10.1016/j.automatica.2023.111123}.}}
\author[First]{Rodrigo Aldana-Lopez*} 
\author[First]{Rosario Aragues} 
\author[First]{Carlos Sagues} 

\address[First]{Departamento de Informatica e Ingenieria de Sistemas (DIIS) and Instituto de Investigacion en Ingenieria de Aragon (I3A), 
\\
Universidad de Zaragoza, Zaragoza 50018, Spain.\\
(e-mail: rodrigo.aldana.lopez@gmail.com, raragues@unizar.es, csagues@unizar.es)}

\begin{abstract} 
In robotic systems, perception latency is a term that refers to the computing time measured from the data acquisition to the moment in which perception output is ready to be used to compute control commands. There is a compromise between perception latency, precision for the overall robotic system, and computational resource usage referred to here as the latency-precision trade-off. In this work, we analyze a robot model given by a linear system, a zero-order hold controller, and measurements taken by several perception mode possibilities with different noise levels. We show that the analysis of this system is reduced to studying an equivalent switching system. Our goal is to schedule perception modes such that stability is attained while optimizing a cost function that models the latency-precision trade-off. Our solution framework comprises three main tools: the construction of perception scheduling policy candidates, admissibility verification for policy candidates, and optimal strategies based on admissible policies.
\end{abstract}


\end{frontmatter}
\section{Introduction}

One of the key ingredients to achieving autonomy of mobile robots is the perception task, which consists of using sensors to obtain an estimate of the state of the robot and the environment. Then, the state of the system can be used to close a feedback loop in order to make the robot follow the desired behavior \cite{siegwart2004}. The term perception latency has been used in \cite{falanga2019} to refer to the perception computing time, measured from the data acquisition to the moment in which perception output is ready to be used to compute control commands. In general, using high-resolution images, a large number of features, and robust feature extraction methods will result in a better perception quality at the expense of larger computing times. \RevOne{This trade-off has been verified experimentally for visual odometry, localization, and mapping in \cite{tcst_codesign,auto_rev11a}. \hlc{In addition, it has been shown that deep learning models used for perception share this type of trade-off as well \cite{scaramuzza2017}}}. However, increasing the perception latency might be undesired since long sampling intervals can prevent the control commands from stabilizing the system \cite{tcst_codesign}. \hlc{In contrast, a smaller latency may be harmful since low-quality measurements lead to faster sampling, which can negatively impact the performance from energetic or CPU load points of view.}

Hence, there is a practical compromise between perception latency and the performance of the overall robotic system, from now on referred to as the latency-precision trade-off.  
The most basic approach in this setting is the following: given a maximum sampling time $T_s$ driving a stable behavior to the robot, use the largest perception latency possible \cite{tcst_codesign}. In other words, \RevThree{$T_s$ poses a real-time} constraint on the perception task of the robot. This approach has been used repeatedly in the robotics literature, e.g. for multi-rotor control using onboard vision sensors \cite{heng,helen,aldana2021}. As studied here, a different approach to overcoming the latency-precision trade-off involves using different perception modes at different moments.

{In this work, we analyze a robot model given by a linear system, a zero-order hold controller, and \RevThree{measurements that can be taken} using several perception mode possibilities with different noise levels. The perception latencies chosen through time dictate the sampling instants for perception and control. We show that the analysis of this system is reduced to studying an equivalent switching system. Moreover, we construct a cost function that includes a control precision term as well as a perception latency penalty which can be used to incorporate the energetic and CPU load points of view. Our goal is to schedule perception modes such that {asymptotic stability for the expected value of the state of the system} is attained while optimizing for the proposed cost function. Our solution framework is composed of three main tools. First, we propose a class of perception scheduling policy candidates for {optimizing the cost of interest}. Second, we devise a procedure for checking if policy candidates are admissible from a stability point of view. Third, we propose and study the performance of optimal strategies based on previously constructed scheduling policies.
\subsection{Related work}
\label{sec:related}
{Scheduling between different sensor configurations was studied in \cite{tcst_sch_sensors} for linear control systems. However, the trade-off between measurement quality and latency was not considered.} This trade-off has been studied for state estimation in \cite{luca2020}. \RevOne{In addition, \cite{rev1b} studied another related type of latency trade-off arising from the quality of a communication channel in networked systems}. Nonetheless, these approaches are mainly focused on the resulting estimation quality rather than the overall performance of a closed-loop system using such an estimation framework. {In \cite{tcst_codesign}, a control-estimator co-design is proposed in a periodic setting, where the relationship between latency and estimator quality is modeled and taken into account in a Model Predictive Control (MPC) strategy. Thus, more emphasis is given to the feasibility of an optimization program rather than on ensuring some form of stability outside the sliding window used in the MPC, which is particularly important when switching between different perception modes.}

\RevOne{The problem of perception latency scheduling is quite close to the notion of scheduling sampling instants for which some works have used event-triggered and self-triggered sampling \cite{Anta2010}.} In these sampling schemes, a state-dependent triggering rule for sampling is obtained by employing sufficient stability conditions. {In particular, \cite{tcst_eventt_sch_info} \RevThree{studies} variable sampling intervals to model information exchange instants in networked control systems and schedule them accordingly using an event-triggered approach. \RevOne{However, these methods do not use a latency model in which shorter sampling intervals lead to poor state estimates as in the latency-precision trade-off.}}

Another approach is to model the effect of variable sampling intervals as a time-varying delay \cite{hetel2017}. However, these works mostly study robustness against arbitrary sampling sequences. On the other hand, as mentioned before, variable sampling problems can be studied using switching systems theory. Switching systems stability results may be divided into two categories. The first one studies stability under arbitrary switching signals. Examples of this type of analysis can be found in \cite{liberzon2003,continous_robust2} for continuous-time and in \cite{rev1a,discrete_robust3} for discrete-time. The switching signal can be considered a state-dependent input in the second category. Examples of the study of stability under state-dependant switching signals can be found in \cite{continous_design1,discrete_continous_design} for continuous-time systems and in \cite{duarte,jungers2014,discrete_design2,discrete_continous_design,mezler} for discrete-time systems. However, as we show later, the connection between stability and optimality may not be evident, particularly with the cost function considered in this work. To this regard, the works \cite{wu,duarte} deal with the optimality of state-dependent switching signals for some particular forms of cost functions that do not extend to the cost function modeling the latency-precision trade-off as we propose.}

\subsection{Notation}
Through this work, $\mf{x}(t)\in\mathbb{R}^n$ represents a continuous-time signal evaluation at $t\in\mathbb{R}$. Moreover, $\mf{x}[k]:=\mf{x}(\tau_k)$ stands for the discrete-time signal obtained by evaluating $\mf{x}(t)$ under a particular sequence of time instants $\{\tau_k\}_{k=0}^\infty$ with $\tau_k\in{\mathbb{R}}$ and $\tau_k<\tau_{k+1}$. Furthermore, $\tr(\bullet)$ stands for the trace of a matrix. Let $\mf{z}\in\mathbb{R}^n$, then $\diag(\mf{z})\in\mathbb{R}^{n\times n}$ represents the diagonal matrix with the components of $\mf{z}$ as the diagonal elements. The symbol $\mathcal{N}(\bar{\mf{x}},P)$ stands for a multi-variate Gaussian distribution with mean $\bar{\mf{x}}\in\mathbb{R}^{n}$ and covariance matrix $P\in\mathbb{R}^{n\times n}$. For any finite set $S$ we denote its cardinality with $|S|$, \RevSeven{and for any finite sequence $a=\{a_i\}_{i=1}^m$, we denote with $\len(a)=m$ its length}. Let $\partial \cset$ and $\interior(\cset) = \cset\setminus\partial \cset$ be the boundary and the interior operators for any set $\cset\subset \mathbb{R}^n$. \RevSeven{In addition, given a finite set $\Gamma$, we usualy denote with $\Gamma'\subseteq\Gamma$ an arbitrary element of the power set $\mathcal{P}(\Gamma)$.}

\section{Problem statement}
\label{sec:problem}

\label{sec:prob}

Consider a mobile robot described by the model
\begin{align}
\begin{split}
    &\nd{\mf{x}}(t) = \left(A\mf{x}(t) + B\mf{u}[k]\right)\nd t + \nd \mf{w}(t)
\label{eq:system}
\end{split}
\end{align}
for intervals of the form $ t\in[\tau_k,\tau_{k+1})$ where $\mf{x}(t)\in\mathbb{R}^n$ is the robot state (e.g. position and velocity for a particle robot),  $\mf{u}[k]\in\mathbb{R}^{n_u}$ is a zero-order hold input (e.g. representing the actuator forces induced to the system by the robot) and a sequence of instants $\tau=\{\tau_k\}_{k=0}^{\infty}$ with $\tau_k\in\mathbb{R}$, $\tau_{k}<\tau_{k+1}$ and $\tau_0\equiv 0$. In addition, $\mf{w}(t)$ is an $n$-dimensional stochastic process capturing disturbances and non-modeled dynamics. 
\begin{assum} 
\label{as:dist} The initial condition of \eqref{eq:system} satisfies $\mf{x}(0)\sim\mathcal{N}(\bar{\mf{x}}_0,P_0)$. Moreover, $\mf{w}(t)$ is a Wiener process with covariance function $\mathbb{E}\{\mf{w}(s)\mf{w}(r)^T\} = W_0\min(s,r)$ and $W_0$ positive semi-definite \cite[Page 63]{astrom}.
\end{assum}

We are interested in studying the error dynamics of the actual robot state with respect to a desired behavior. Hence, to simplify the exposition, $\mf{x}(t)$ will represent such error which is desired to be maintained as close as possible to the origin. However, motivated by the perception latency problem, we consider that at $t=\tau_k$ a state estimate cannot be obtained directly. Instead, indirect information is obtained (e.g. an image captured by a camera) that has to be processed through a perception procedure. To do so, there are $D$ available perception modes where the quality of their estimates depends on the perception latency $\Delta^1,\dots,\Delta^D$ for each perception mode.
\begin{assum}
\label{as:meas}
Let $C\in\mathbb{R}^{n_z\times n}$ be the measurement matrix for the perception setting with the pair $(A,C)$ observable and $p_k\in\{1,\dots,D\}$ be the perception mode chosen at $t=\tau_k$. Then, the output of the perception method is $\mf{z}[k] = C\mf{x}[k]+\mf{n}[k]$, available at $t=\tau_k+\Delta^{p_k}$, with noise $\mf{n}[k]\sim\mathcal{N}(0,\Sigma^{p_k})$. Here, $\Sigma^1,\dots,\Sigma^D$ capture the different noise levels of the accuracy of methods $1,\dots,D$. 
\end{assum}

\RevOne{The setting in which different perception modes are available, achieving different noise levels and perception latency, captures the actual practical behavior of usual perception algorithms 
when different configurations are available \cite{luca2020,tcst_codesign,falanga2019,auto_rev11a}. The values of the latency $\Delta^{p_k}$ and noise level $\Sigma^{p_k}$ for a concrete perception method $p_k$ can be obtained offline through a statistical analysis of its performance. In practice, it is expected that a longer latency $\Delta^{p_k}$ results in a more concentrated distribution $\mathcal{N}(0,\Sigma^{p_k})$. For instance, in \cite{luca2020}, $\Sigma^{p_k}$ is modeled to be inversely proportional to the latency, which is also supported by the experimental data in \cite{tcst_codesign,auto_rev11a}. However, to allow our methods to be tailored to 
any latency-precision model, we do not assume a particular relation between the latency $\Delta^{p_k}$ and its corresponding noise level $\Sigma^{p_k}$ but that these might be different between perception methods.}

Due to  Assumption \ref{as:meas} the only available information at $t=\tau_k$ is $\{\mf{z}[0],\dots,\mf{z}[k-1]\}$. Hence, a control is designed as $\mf{u}[k]=L^{p_k}\hat{\mf{x}}[k|k-1]$ where $\hat{\mf{x}}[k|k-1] = \mathbb{E}\left\{\mf{x}[k]\ \big |\ \mf{z}[0],\dots,\mf{z}[k-1]\right\}$ is the conditional mean of $\mf{x}[k]$ using old measurements as given in Appendix \ref{ap:evol}. \RevTen{Moreover, the gains $L^1,\dots,L^D$ are assumed to be given, which may have been designed for each latency by separate, provided that the pair $(A,B)$ is controllable.} Furthermore, note that at $t=\tau_k+\Delta^{p_k}$ the robot is ready to take a new measurement, thus we set $\tau_{k+1}=\tau_k+\Delta^{p_k}$. Finally, any (perhaps infinite) sequence of perception modes $p$ is referred to as a \textit{perception schedule}. We refer to a rule that generates a perception schedule $p$ based on state information as a \textit{scheduling policy}.

In this context, the goal of this work is the following:

\begin{prob}[Perception Scheduling Problem]
\label{prob:main}
Consider an interval of interest $[0,T_f]$ and $\textsf{att}(p;[0,T_f])$, referred to as the \emph{attention} of $p$, the amount of sampling instants in the interval $[0,T_f]$ induced by a perception schedule $p$. Moreover, let $0<r^{p_k}$ be the associated penalty for each perception mode. Thus, given Assumptions \ref{as:dist} and \ref{as:meas}, and given $\lambda_\mf{x},\lambda_r,T_f>0$  and $Q,Q_f$ positive semi-definite, the problem is to find a perception schedule $p=\{p_k\}_{k=0}^\infty$ such that

\begin{equation}
\begin{aligned}
    \label{eq:cost_functional}
    &\mathcal{J}(p;\mf{\bar{x}}_0,P_0) =  \frac{\lambda_r}{T_f} \sum_{k=0}^{\alpha-1} r^{p_k}\\&+  {\lambda_{\mf x}}\mathbb{E}\left\{\frac{1}{T_f}\int_{0}^{T_f} \mf{x}(t)^TQ\mf{x}(t)\ \nd t + \mf{x}(T_f)^TQ_f\mf{x}(T_f) \right\}
    \end{aligned}
 \end{equation}
 with $\alpha=\textsf{att}(p;[0,T_f])$, is minimized for system \eqref{eq:system}. Moreover, if there exists a stabilizing perception schedule for system \eqref{eq:system} under arbitrary $T_f$, then $p$ must induce $\lim\limits_{t\to\infty}\mathbb{E}\{\mf{x}(t)\}=0$.
\end{prob}

\RevThree{As usual in minimum variance control problems \cite[Page 172]{astrom} the persistent introduction of uncertainty originated from disturbances in the model, and measurement noise prevents a linear controller from making $\mf{x}(t)$ converge to the origin in the mean-squared error sense. Hence, to concentrate our efforts on building a schedule $p$, it is more practical to ensure convergence of $\mathbb{E}\{\mf{x}(t)\}$ towards the origin, and deal with the effect of the second order moment of $\mf{x}(t)$ during the finite interval of interest $[0,T_f]$ by minimizing \eqref{eq:cost_functional}.}

The cost in \eqref{eq:cost_functional} is meant to model the latency-precision trade-off we described before in the following way. The second term in \eqref{eq:cost_functional} is composed of an expected quadratic penalty for cost $\mf{x}(t)$ of \eqref{eq:system}, which penalizes deviations from the origin. On the other hand, the first term in \eqref{eq:cost_functional} is the accumulation of the penalties $r^{p_0},r^{p_1},\dots$ over the interval $[0,T_f]$. If $r^{p_0}=r^{p_1}=\cdots=1$, this term is proportional to the attention of $p$. Keeping the number of sampling events small is desirable both from an energetic point of view and to minimize the use of I/O buses when using the sensors. On the other hand, consider that the latency $\Delta^{p_k}$ is not used exclusively for perception, but also that the method frees the computing unit for an interval of length $(1-f^{p_k})\Delta^{p_k}$ with $f^{p_k}\in(0,1)$. Thus, the \textit{perception CPU load} in the interval $[0,T_f]$ is $(1/T_f)\sum_{k=0}^{\alpha-1}f^{p_k}\Delta^{p_k}$ and can be taken into account in \eqref{eq:cost_functional} by using $r^{p_k} = f^{p_k}\Delta^{p_k}$.

Consider the last requirement in Problem \ref{prob:main}. Note that the cost in \eqref{eq:cost_functional} only penalizes the schedule during the interval $[0,T_f]$, which can be used to penalize a transient response. In practice, it may be beneficial to pose a similar problem once $[0,T_f]$ has elapsed in a moving horizon fashion. However, this strategy won't {ensure asymptotic stability for $\mathbb{E}\{\mf{x}(t)\}$} on the long run for arbitrary values of $\lambda_{\mf{x}},\lambda_r,T_f,Q,Q_f$ since the individual terms in \eqref{eq:cost_functional} are often conflicting. As an example, take $\lambda_{\mf{x}}=0$ and $\lambda_r>0$ in which the system's error is not penalized and thus, minimizing \eqref{eq:cost_functional} won't ensure stability regardless of $T_f$ or how often the optimization problem is solved. Hence, a connection between optimality and stability is not evident in general problem setting. 

Our strategy is based on the following observation. Using Proposition \ref{prop:distro} in Appendix \ref{ap:evol}, it is obtained that $\bar{\mf{x}}[k]:=\mathbb{E}\{\mf{x}[k]\}$ is given by
\begin{equation}
\label{eq:switch}
\mf{\bar{x}}[k+1]=\Lambda(\Delta^{p_k})\mf{\bar{x}}[k], \ \ \mf{\bar{x}}[0]=\bar{\mf{x}}_0
\end{equation}
where $\Lambda(\Delta^{p_k}) := \exp(A\Delta^{p_k})+ \int_0^{\Delta^{p_k}}\exp(A\tau)\nd \tau BL^{p_k}$. As a result, \eqref{eq:switch} is a switched system which switches between matrices $\{\Lambda(\Delta^{1}),\dots,\Lambda(\Delta^{D})\}$ according to the perception schedule $p$ as the switching signal. Hence, the {asymptotic stability} character of Problem \ref{prob:main} is tied to the stability of the switching system in \eqref{eq:switch}. In this context, the outline of our solution has 3 main components:
\begin{itemize}
    \item\RevOne{ First, in Section \ref{sec:stability} we analyze some results in the literature regarding stabilizing switching signals for discrete-time systems. We provide an extension of the results found in the literature to enable the construction of multiple scheduling policy candidates {for optimizing \eqref{eq:cost_functional}}. These scheduling policies are all stabilizing for \eqref{eq:switch} when an admissibility condition is attained.
    \item Second, for completeness, in Section \ref{sec:admissible} }\RevOne{we provide a new algorithm to check if a scheduling policy candidate is admissible.}  
    \item \RevOne{Third, in Section \ref{sec:balance} we provide a new sub-optimal algorithm for Problem \ref{prob:main} based on the multiple scheduling policies previously constructed. We analyze its theoretical performance and discuss some heuristics and approximations.}
\end{itemize}
\section{Stability preserving perception schedules}
\label{sec:stability}
In this section, we develop scheduling policies which ensure \eqref{eq:switch} is stable. 
In \cite{jungers2014} an interesting framework is proposed which provides sufficient and necessary conditions for {asymptotic stability} of discrete-time switching linear systems as \eqref{eq:switch}, and generates state-dependant switching laws based on such conditions. In the following, we analyze some related ideas and propose a framework suitable for Problem \ref{prob:main}. This is required since not only a single switching rule is needed, but several switching rule possibilities have to be available as the search space for the optimization of \eqref{eq:cost_functional}. First, let us introduce the following concept:
\begin{defn}
A set of schedules is any set of the form $\Gamma = \{\gamma^1,\dots,\gamma^{|\Gamma|}\}$ with $|\Gamma|<\infty$ where $\gamma^i\in\{1,\dots,D\}^{\len(\gamma^i)}$ are individual schedules with $\len(\gamma^i)<\infty, \ \forall i\in\{1,\dots,|\Gamma|\}$.
\end{defn}

We aim to construct stabilizing perception schedules by piecing together individual finite-length schedules found in a set of schedules. Up to this point, sets of schedules are arbitrary and can be constructed randomly. However, stabilizing switching signals cannot arise for any set of schedules. In the following, we study which sets of schedules induce stability. Consider the following objects: the hyper-ellipse
$$
\cset_0 := \{\mf{x}\in\mathbb{R}^n :\mf{x}^TM_0\mf{x} \leq 1\}
$$
for some arbitrary positive definite matrix $M_0$ and
$$
\cset_\seq := \{\mf{x}\in\mathbb{R}^n:\mf{x} =\Lambda(\Delta^{\seq_0})^{-1}\cdots \Lambda(\Delta^{\seq_{\ell-1}})^{-1}\mf{y},  \mf{y}\in\cset_0\}
$$
where $\seq:=\{\seq_k\}_{k=0}^{\ell-1}$ is a schedule of finite length $\len(\seq)=\ell$. Thus, by construction, for any $\mfb{x}_0\in\cset_\seq$, then $\mfb{x}[\len(\seq)]\in\cset_0$ for such schedule $\seq$. This is verified as:
\begin{equation*}
\begin{aligned}
&\mfb{x}[\len(\seq)] = \Lambda^\seq \mfb{x}_0 \in \Lambda^\seq \cset_{\seq}\\
&= \{\mf{z}\in\mathbb{R}^n:\mf{z} = \Lambda^\seq \mf{x}\andset \mf{x} = (\Lambda^\seq)^{-1} \mf{y}\andset \mf{y}^TM_0\mf{y} \leq 1\} \\
&=\{\mf{z}\in\mathbb{R}^n:\mf{z} = \Lambda^\seq (\Lambda^\seq )^{-1}
\mf{y}\andset \mf{y}^TM_0\mf{y} \leq 1\} \\
&=\{\mf{z}\in\mathbb{R}^n: \mf{z}^TM_0\mf{z} \leq 1\} = \cset_0
\end{aligned}
\end{equation*}
with $\Lambda^\seq:=\Lambda(\Delta^{\seq_{\ell-1}})\cdots \Lambda(\Delta^{\seq_0})$ as the multiplication chain in \eqref{eq:switch} for the schedule $\seq$. Moreover, note that $\cset_\seq$ 
has the shape of an hyper-ellipse defined by the positive definite matrix $M_\seq = (\Lambda^\seq)^TM_0(\Lambda^\seq)$ as $\cset_\seq \equiv \{\mf{x}\in\mathbb{R}^n:\mf{x}^TM_\seq\mf{x} \leq 1\}$. 

The key idea is the following. Consider $\ses$ to be a set of schedules $\seq$. Thus, if the interior of the set 
\begin{equation}
\label{eq:cset}
\cset_\ses^\sstar := \bigcup_{\seq\in\ses}\cset_\seq
\end{equation}
contains $\cset_0$, it means that for each initial condition, there exists a schedule $\seq\in\ses$ which contracts $\cset_\ses^\sstar$ into itself after $\len(\seq)$ steps. This is, if $\mfb{x}_0\in\cset_\ses^\sstar$, thus $\mfb{x}[\len(\seq)]\in\cset_0\subset\interior(\cset_\ses^\sstar)$ for some $\seq\in\ses$. 
\begin{defn}[Admissible set of schedules] A finite set of schedules $\ses$ is admissible if $\cset_0\subset \interior(\cset_\ses^\sstar)$.
\end{defn}
As we show in a subsequent result, if $\Gamma$ is admissible, a contracting schedule can be found $\forall \bar{\mf{x}}\in\mathbb{R}^{n}$ as:
\begin{equation}
\label{eq:sw_law}
\seq^\sstar(\mfb{x};\ses) := \underset{\seq\in\ses}{\text{arg min}} \{ r\in\mathbb{R} : r = \mfb{x}^TM_\seq\mfb{x}\}
\end{equation}
Using this switching rule for various admissible sets of schedules $\Gamma^1,\dots,\Gamma^m$, the following strategy can be used online to compute a stabilizing perception method at each $t=\tau_k$ given $\mf{\bar{x}}[k]$.
\begin{defn}
\label{def:sps}[Stability Preserving Scheduling] The $\sptwo$ (Stability Preserving Scheduling Policy) strategy is defined by Algorithm \ref{algo:sp2}.
\end{defn}

\begin{algorithm}
 \caption{$\sptwo$ strategy}
 \begin{algorithmic}[1]
 \renewcommand{\algorithmicrequire}{\textbf{Input:}}
 \renewcommand{\algorithmicensure}{\textbf{Output:}}
 \REQUIRE $\mfb{x}_0$.\\
 \ENSURE  Perception schedule $p$.\\
  \STATE $\seq\leftarrow\varnothing$.
  \STATE $i\leftarrow0$
  \FOR { each instant $\tau_k, k=0,1,\dots$}

  \IF {($i < \len(\seq)$)} 
  \STATE $p_k\leftarrow \seq_i$ \RevThree{\#{For chosen schedule $\gamma$, traverse its elements $\gamma_0,\dots,\gamma_{\len(\gamma)}$}.}
  \STATE $i\leftarrow i+1$
  \ELSE  
  \STATE $\ses\leftarrow$ any set of schedules chosen from the admissible options $\Gamma^1,\dots,\Gamma^m$.
  \STATE $\seq\leftarrow\seq^\sstar(\mfb{x}[k];\ses)$ from \eqref{eq:sw_law}. \RevThree{\#{Once previous schedule has been consumed, choose a new schedule}}
  \STATE $i\leftarrow 0$
  \ENDIF
  \ENDFOR
 \end{algorithmic}
 \label{algo:sp2}
 \end{algorithm}
\begin{algorithm} 
 \caption{Construction of a set of schedules}
 \begin{algorithmic}[1]
 \renewcommand{\algorithmicrequire}{\textbf{Input:}}
 \renewcommand{\algorithmicensure}{\textbf{Output:}}
 \REQUIRE $\ell,D$.\\
 \ENSURE  $\Gamma$.\\
 \STATE $\Gamma = \varnothing$
  \REPEAT
  \IF {($\{1,\dots,D\}^{\ell}\setminus \Gamma = \varnothing$)}
  \STATE \# If all schedules have been evaluated
  \STATE Increase $\ell$
  \ENDIF
  \STATE \# Generate a random sequence over $\{1,\dots,D\}$ with random length $\ell'\leq \ell$.
  \STATE $\ell'\leftarrow \texttt{randomSample}(\{1,\dots,\ell\})$ 
  \STATE $\Gamma \leftarrow \Gamma \bigcup \texttt{randomSample}(\{1,\dots,D\}^{\ell'}\setminus \Gamma)$ 

  \UNTIL{$\mathbb{S}_0\subset\interior(\mathbb{S}_\Gamma^\sstar)$ \# Using Algorithm \ref{algo:admisibility}}

 \end{algorithmic}
 \label{algo:ses}
 \end{algorithm}
\begin{rem} 
\label{rem:difference}
A similar rule to \eqref{eq:sw_law} was studied in \cite{jungers2014} for a particular fixed set of schedules $\Gamma$ constructed {through} a combinatorial approach. This is, all possible combinations of elements in $\{1,\dots,D\}$ up to the first length at which admissibility is attained \cite[Algorithm 1]{jungers2014}. \RevEleven{As a result, a single scheduling policy is considered regardless of the choice of \eqref{eq:cost_functional}, and with no additional degrees of freedom to improve performance. Unlike the previous approach, we introduce an extra degree of freedom by constructing multiple sets of schedules $\Gamma^1,\dots,\Gamma^m$. Then, as explained in detail in Section \ref{sec:balance}, a switching law of the form \eqref{eq:sw_law} is used for each set of schedules as scheduling policy candidates to optimize for \eqref{eq:cost_functional}. This makes our approach more beneficial for the perception scheduling problem since intuitively, given an initial condition $\mfb{x}[0]$, the approach with a single set of schedules will result in a fixed schedule $p$, regardless of the cost \eqref{eq:cost_functional} whereas our method can adapt $p$ to the cost.}
\end{rem}

In this sense, an alternative for the combinatorial approach is given in Algorithm \ref{algo:ses} in which random schedules are appended to $\Gamma$. These schedules are generated with length up to a given value $\ell$, unless all schedules are contained in $\Gamma$, in which case $\ell$ is increased. This procedure terminates once $\Gamma$ is admissible and, similarly to the combinatorial approach, this must happen if there \RevThree{exists} a stabilizing schedule for \eqref{eq:switch}. When compared to the combinatorial approach, the advantage of using Algorithm \ref{algo:ses} is two-fold. First, depending on the initial value of the length $\ell$, longer sequences can be tested earlier. This is complicated using the combinatorial approach due to the combinatorial explosion involved in such a strategy. \RevThree{Note that even if the combinatorial approach is used starting from an initial length $\ell>1$ to test longer sequences earlier, this strategy will neglect short sequences, which may also be useful for rapid decision-making if available. Hence, Algorithm \ref{algo:ses} covers short and long sequences from the beginning. }Moreover, Algorithm \ref{algo:ses} allows us to generate multiple admissible sets of schedules, namely $\Gamma^1,\dots,\Gamma^m$ useful for our optimization strategy. 

However, since in line 8 of Algorithm \ref{algo:sp2} we allow to change the set of schedules $\Gamma$ to any admissible one between the $m$ options $\{\Gamma^1,\dots,\Gamma^m\}$, the scheduling policy in \eqref{eq:sw_law} may be different each time in line 9 of Algorithm \ref{algo:sp2} is executed. The following results show that even in this case, Algorithm \ref{algo:sp2} manages to stabilize \eqref{eq:switch}.

\begin{thm}
\label{th:stability}
Assume that there exists at least one admissible set of schedules and that the $\sptwo$ strategy is used in \eqref{eq:switch}. Thus $\lim_{k\to\infty}\mfb{x}[k] = 0$.
\end{thm}

\begin{pf}
All proofs are provided in Appendix \ref{sec:proofs}.
\end{pf}

Checking if a set of schedules is admissible is not a trivial task. \cite[Remark 6]{jungers2014} proposes to check if there is at least one $\gamma\in\Gamma$ such that $\mathbb{S}_0\subset\interior(\cset_\seq)$ as a sufficient condition for admissibility. However, this is a very conservative condition that won't be attained in general, even when $\Gamma$ is admissible. Motivated by this, in the following section, we provide a novel analysis to check if a set of schedules is admissible.
\section{Non-conservative admissibility checking}
\label{sec:admissible}
The basis of a non-conservative admissibility checking is formulated as a nonlinear program as follows.
\begin{thm}
\label{eq:optim_theo}
Let 
\begin{equation}
\begin{aligned}
\label{eq:program}
R &:= \min_{\mfb{x}\in\partial \cset_\ses^\sstar} \mfb{x}^TM_0\mfb{x}.
\end{aligned}
\end{equation}
Then, $\cset_0\subset \cset_\ses^\sstar$ if and only if $R>1$.
\end{thm}

Hence, under the light of Theorem \ref{eq:optim_theo}, studying if $\ses$ is admissible is equivalent to solving the nonlinear program in \eqref{eq:program} and checking if $R>1$. {Note that the attempt to solve numerically for local minima in \eqref{eq:program} may fail since the condition $R>1$ is only useful when the global minimum is considered. To the best of our knowledge, there \RevThree{is no previous work which solves for the global minimum} of a non-convex program of the form \eqref{eq:program}. Thus, the following efforts are dedicated to this task.}

Note that most of the theory presented up to now remains the same even when $\mathbb{S}_0$ is not a hyper-ellipse. An arbitrary $C^\sstar$ set as in Definition \ref{def:cstar} in Appendix \ref{ap:cstar} can be used instead of a hyper-ellipse. For example, polyhedral sets are also considered in \cite{jungers2014}. However, note that the set $\partial\cset_\ses^\sstar$ may be non-convex, implying that there might be many local minima for \eqref{eq:program}, which complicates the analysis. Despite this, we show that for the case where $\mathbb{S}_0$ is a hyper-ellipse, we can compute the global minimum of \eqref{eq:program}. The strategy is the following: we  split the nonlinear program in \eqref{eq:program} into subprograms of the following form:
\begin{equation}
    \begin{aligned}
    \label{eq:subproblem}
&\min_{\mfb{x}\in\mathbb{R}^n}\mfb{x}^TM_0\mfb{x} \\
\text{s.t. :}&\  h_\seq(\mfb{x}):=\mfb{x}^TM_\seq\mfb{x}-1 = 0, \ \ \RevThree{\forall\seq\in{\ses}'\in\mathcal{P}({\ses})}
    \end{aligned}
\end{equation}
\RevThree{Where $\ses'$ is any subset of $\ses$, i.e. $\ses'$ is in the power set $\mathcal{P}(\ses)$. Intuitively, the sub-program in \eqref{eq:subproblem} is useful since due to the shape of $\partial\mathbb{S}^\sstar_\Gamma$, the global optimum of \eqref{eq:program} must lie either in the boundary of a single hyper ellipse $\mathbb{S}_\gamma$ for some $\gamma\in\Gamma$ which has the form $h_\gamma(\bar{\mf{x}})=0$, or in the intersection of the boundaries of multiple hyper-ellipses. As an example, consider $\Gamma=\{p,q\}$ with arbitrary schedules $p,q$. Hence, the global optimum of \eqref{eq:program} must be the global optimum of \eqref{eq:subproblem} with either $\Gamma'=\{p\}$, $\Gamma'=\{q\}$, which correspond to checking points in $\partial\mathbb{S}_{p},\partial\mathbb{S}_{q}$ by separate or $\Gamma'=\{p,q\}$ which corresponds to checking points in $\partial\mathbb{S}_{p}\cap \partial\mathbb{S}_{q}$ where the two constraints $h_{p}(\bar{\mf{x}})=0, h_{q}(\bar{\mf{x}})=0$ are active. Therefore, the solution for \eqref{eq:program} comes from \eqref{eq:subproblem} for some $\Gamma'\in\mathcal{P}(\Gamma)=\{\varnothing,\{p\},\{q\},\{p,q\}\}$. This idea is formalized in the following:}
\begin{cor}
\label{le:subproblem}
\RevSeven{Let $\mfb{X}_{\ses'}$ be the set of all critical points (local minima candidates)} $\mfb{x}^*$ of \eqref{eq:subproblem} such that $\mfb{x}^*\notin \cset_\seq$ for any $\seq\in\ses\setminus\ses'$ and any $\ses'\in\mathcal{P}(\ses)$. Thus, the global minimum $R$ of \eqref{eq:program} can be obtained as

\begin{equation}
    \label{eq:subproblem3}
R = \min\left\{r=\mfb{x}^TM_0\mfb{x} : \mfb{x}\in \bigcup_{\ses'\in\mathcal{P}(\ses)}\mfb{X}_{\ses'}\right\}
\end{equation}
\end{cor}

Corollary \ref{le:subproblem} implies that the solution of \eqref{eq:program} can be obtained by solving programs of the form \eqref{eq:subproblem}, where critical points of \eqref{eq:subproblem} are rejected if such points are contained in other hyper-ellipses not involved in \eqref{eq:subproblem} (written as $\mfb{x}^*\notin \cset_\seq,\forall \seq\in\ses\setminus\ses'$ in Corollary \ref{le:subproblem}). Moreover, all programs of the form \eqref{eq:subproblem} involve from one hyper-ellipse at the time, to all combinations of $\cset_\seq, \forall \seq\in\ses$.

\begin{rem}
\label{rem:homotopy}
Note that if $|\ses'|=n$ in \eqref{eq:subproblem}, the set of all points $\mfb{x}$ with $\mfb{x}^TM_\seq\mfb{x}=1,\forall\seq\in\ses'$ contains only isolated points for almost all $\{M_\seq\}_{\seq\in\ses}$. Moreover, the set of equations $\mfb{x}^TM_\seq\mfb{x}=1$ are $n$ polynomial equations with $n$ variables which have $n^n$ complex solutions due to the Bezout's Theorem \cite[Theorem 4.14]{grobner}. Homotopy solvers such as the one in \cite{phc} can track all such real solutions for polynomial systems or obtained explicitly if $n\leq 2$. Hence, all reals solutions can be checked directly to build the set $\mfb{X}_{\ses'}$. In contrast, the case with $|\Gamma'|>n$ can be ignored since those solutions are either contained in the $|\Gamma'|=n$ case or are nonexistent.
\end{rem}

\RevTen{In general, an optimization program of the form \eqref{eq:subproblem} can be solved by the Lagrange multipliers method. However, depending on the problem constraints, in this case, given in the form $h_\gamma(\bar{\mf{x}})=0$, not all local minima can be found through this method. Such local minima are often called non-regular \cite[Page 279]{nonlinear_prog}. Due to the fact that we look for the global optimum of \eqref{eq:subproblem}, we are forced to analyze both regular and non-regular points. Formally, critical points in $\mfb{X}_{\ses'}$ for  \eqref{eq:subproblem} in which $|\ses'|<n$ will be classified into regular or non-regular points for their analysis as follows:}
\begin{defn}\cite[Page 279]{nonlinear_prog}\label{def:regular}
A critical point $\bar{\mf{x}}$ for \eqref{eq:subproblem} is said to be regular if the vectors $\nabla h_\seq(\mfb{x}) = 2M_\seq\mfb{x}, \seq\in\ses'$ are linearly independent.
\end{defn}
\RevTen{In the next section, we use the Lagrange multiplier method to find the list of critical regular points for \eqref{eq:subproblem}. Then, we use a tailored analysis for the non-regular ones.}
\subsection{Regular points analysis}

We start the study of \eqref{eq:subproblem} with the characterization of regular points by means of the Lagrange multiplier theorem in \cite[Proposition 3.1.1]{nonlinear_prog}.
\begin{thm}
\label{th:regular}
Let ${\lambda}:=\{\lambda_{\seq}\}_{\seq\in \ses'}$ be a solution to the system of equations
\begin{subequations}
\label{eq:poly_eq}
\begin{align}
\text{\normalfont tr}(G(\lambda)^\dagger(M_\seq-M_\seqq))&=0,\ \ \seq,\seqq\in\ses' \label{eq:traces}\\
\text{\normalfont det}\ G(\lambda)&=0 \label{eq:det_cond}
\end{align}
\end{subequations}
where 
\begin{equation}
\label{eq:Gmat}
G(\lambda) := M_0 + \sum_{\seq\in\ses'} \lambda_\seq M_\seq
\end{equation}
and $G(\lambda)^\dagger$ is the adjugate matrix of $G(\lambda)$ \cite[Page 22]{horn}. Thus, all regular critical points $\mfb{x}^*$ of program \eqref{eq:subproblem} are  in the kernel of $G(\lambda)$ for some $\lambda$. This is, there exists a solution $\lambda$ of \eqref{eq:poly_eq}  such that $G(\lambda)\mfb{x}^*=0$ and have $(\mfb{x}^*)^TM_0(\mfb{x}^*) = -\sum_{\seq\in\ses'}\lambda_\seq$.
\end{thm}

\begin{rem}
\label{rem:tracedet}
Note that the number of equations in \eqref{eq:poly_eq} are evidently more than $|\ses'|$. However, most of them are linearly dependent. They can be arranged as $|\ses'|$ independent equations with $|\ses'|$ variables by using $|\ses'|-1$ equations of the form \eqref{eq:traces} in addition to \eqref{eq:det_cond}. These $|\ses'|$ equations are independent for most choices of $\{M_\seq\}_{\seq\in\ses}$. Moreover, note that all equations in \eqref{eq:poly_eq} are polynomials in $\lambda$ and therefore, the number of solutions can be counted by means of Bezout's theorem and obtained similarly as pointed out in Remark \ref{rem:homotopy} using the algorithm in \cite{phc}.
\end{rem}

Moreover, the nullity (dimension of kernel) of the singular matrix $G(\lambda)$ may be of dimension up to $n$. However, among the possible singular matrices, ones with nullity greater than one lie in a set of measure zero (with respect to the possible choices of $\{M_\seq\}_{\seq\in\ses}$). This fact is evidenced in the following result.

\begin{cor}
\label{cor:nullity}
Let $\lambda$ satisfy \eqref{eq:poly_eq}, assume that $G(\lambda)$ has nullity greater than 1, and let $G_{ij}(\lambda)$ be the sub-matrix obtained by deleting the $i$-th row and $j$-th column of $G(\lambda)$.
Then, $\lambda$ complies with 
\begin{equation}
    \label{eq:sub}
    \text{\normalfont det }G_{ij}(\lambda) = 0,\ \  \forall(i,j)\in \{1,\dots n\}^2
\end{equation}
\end{cor}
As a result of Corollary \ref{cor:nullity}, the case in which $G(\lambda)$ has nullity greater than \RevThree{one is} is not common in a general setting since $\lambda\in\mathbb{R}^{|\ses'|}$ would have to comply $|\ses'|-1$ equations of the type \eqref{eq:traces} in addition to $n^2$ equations of the type \eqref{eq:sub}. Henceforth, we ignore such cases in our analysis. 
Now, for $G(\lambda)$ with nullity 1, let ${\mf{g}}$ be any vector in the kernel of $G(\lambda)$. Thus, as a result of Theorem \ref{th:regular} the critical regular points for such $\lambda$ are parallel to $\mf{g}$, and can be uniquely computed by scaling $\mf{g}$ in order to comply $(\mfb{x}^*)^TM_0\mfb{x}^* = -\sum_{\seq\in\ses'}\lambda_\seq$ as:
\begin{equation}
\label{eq:null_space}
\mfb{x}^*= \pm{\mf{g}}\sqrt{\frac{-\sum_{\seq\in\ses'}\lambda_\seq}{{\mf{g}}^TM_0{\mf{g}}}}
\end{equation}
and therefore, the condition $\mfb{x}^*\in\cset_{\seq}$ for $\seq\in\ses\setminus\ses'$ can be checked by computing if $(\mfb{x}^*)^TM_\seq(\mfb{x}^*)>1$ in order to build the set $\mfb{X}_{\ses'}$ as required in Corollary \ref{le:subproblem}.

\subsection{Non-regular points analysis}
Non-regular points for program \eqref{eq:subproblem} are points $\mfb{x}$ which comply $\mfb{x}^TM_\seq\mfb{x}=1, \forall \seq\in\ses'$ and the vectors $\{M_\seq\mfb{x}\}_{\seq\in\ses'}$ linearly dependent. In general, linearly dependent vectors are rare among all possible vectors, and hence is not surprising that non-regular points won't exist in arbitrary settings. In the following result, we characterize when they exist by obtaining what additional concrete conditions must be satisfied.

\begin{prop}
\label{prop:non_regular}
Let $W(\mfb{x})$ an $n\times |\ses'|$ matrix whose columns are the vectors $M_\seq \mfb{x}, \forall \seq\in\ses'$ and let $W_{\alpha}(\mfb{x})$ be the $|\ses'|\times |\ses'|$ sub-matrix of $W(\mfb{x})$ obtained by deleting $n-|\ses'|$ rows indexed in $\alpha\in\mathcal{A}\subset\{1,\dots n\}^{n-|\ses'|}$ with $\mathcal{A}$ containing all elements in $\{1,\dots n\}^{n-|\ses'|}$ with non repeating entries. Therefore,
any non-regular point for program \eqref{eq:subproblem} must comply with
\begin{subequations}
\label{eq:nr_eq}
\begin{align}
\mfb{x}^TM_\seq \mfb{x}&=1,\ \ \forall\seq\in\ses' \label{eq:nr_constraints}\\
\text{\normalfont det}\ W_\alpha(\mfb{x})&=0,\ \  \forall\alpha\in\mathcal{A} \label{eq:nr_det}
\end{align}
\end{subequations}
\end{prop}
Note that the amount of equations of the type \eqref{eq:nr_det} are equivalent to the number of sub-matrices of size $|\ses'|\times |\ses'|$ inside $W(\mfb{x})$ which is exactly $n!/(|\ses'|!(n-|\ses'|)!)$ \cite[Page 21]{horn}. Note that there are at least $n-|\ses'|+1$ independent equations of this type since that correspond to the number of $|\ses'|\times |\ses'|$ sub-matrices of $W(\mfb{x})$ with consecutive rows. Hence, the number of equations in \eqref{eq:nr_eq} are $|\ses'|$ of the type \eqref{eq:nr_constraints} and at least $n-|\ses'|+1$ of type \eqref{eq:nr_det}. Equivalently, a non-regular point $\mfb{x}\in\mathbb{R}^n$ would have to comply with at least $n+1$ polynomial equations. One could solve $n$ equations picked from \eqref{eq:nr_eq} and check if the results comply with the remaining equations. However, due to this over-determined \RevThree{feature} of equations in \eqref{eq:nr_eq} non-regular points won't exist in general.

\subsection{Admissibility checking algorithm}
Using the results in the previous sections, we summarize the methodology for admissibility checking in the following. Consider Algorithm \ref{algo:admisibility} which takes as an input the dimension of the state space $n$, the matrix $M_0$, the set of matrices $M:=\{M_\seq\}_{\seq\in\ses}$ and the set of schedules $\ses$. This function builds a set $\mathcal{R}$ of all global minima candidates of \eqref{eq:program} and checks $\min\mathcal{R}>1$ as this condition is equivalent to admissibility due to Theorem \ref{eq:optim_theo}. In line 2, all subsets of schedules in $\ses$ are checked, ranging from no schedules at all, one schedule at the time, pairs of schedules, and so on. In line 3 we check if the constraint space is of dimension $n$, since in such case the constraint space is comprised of just isolated points, and we obtain critical points as in Remark \ref{rem:homotopy} by calling \texttt{IsolatedSolutions}. Otherwise, we obtain regular values by calling \texttt{RegularSolutions}. Cases in which the constraint space is of dimension more than $n$ are ignored. Algorithm \ref{algo:measurezero} obtains critical objective values when the constraint space contains only isolated points. This algorithm looks for real solutions of the system of $n$ equations and $n$ variables $\{\mfb{x}^TM_\seq\mfb{x}=1,\forall\seq\in\ses\}$ in line 1. Then, for all  solutions we check one by one if they are not contained in other hyper-ellipses for $\seq\in\ses\setminus\ses'$. If any of those conditions are complied we reject such points in line 3 in order to build the set $\mfb{X}_{\ses'}$ required in Corollary \ref{le:subproblem}. Algorithm \ref{algo:regular} looks for regular points of \eqref{eq:program}. We build a set of critical objective values in $\mathcal{R}$ and then return $r\leftarrow\min\mathcal{R}$. First, we solve the system of equations in \eqref{eq:poly_eq}. Then, for all real solutions $\lambda$ of such polynomial equations we compute $G(\lambda)$ and the critical point as in \eqref{eq:null_space}. These points are rejected if they are contained in other hyper-ellipses $\ses\setminus\ses'$ in line 7. Then, the critical objective is given by $-\sum_{\seq\in\ses'}\lambda_\seq$ as obtained in Theorem \ref{th:regular}.

\begin{rem}
Note that in Algorithm \ref{algo:admisibility} we explicitly assumed that $G(\lambda)$ has nullity 1 and that there are no non-regular points. This assumption is reasonable since  the system of equations made by \eqref{eq:traces} in addition to \eqref{eq:sub} and the system of equations in \eqref{eq:nr_eq} are over-determined for almost any $\{M_\gamma\}_{\gamma\in\Gamma}$. 
\end{rem}
\begin{rem}
Note that due to line 2 in Algorithm \ref{algo:admisibility}, the complexity of the admissibility checking procedure grows as $|\mathcal{P}(\Gamma)| = 2^{|\Gamma|}$. Still, this procedure can be stopped as soon as some value contained $\mathcal{R}$ is less than 1. Moreover, recall that checking for admissibility is not required online for the $\sptwo$ strategy in Algorithm \ref{algo:sp2}. Instead, this procedure is used to construct admissible sets of schedules in Algorithm \ref{algo:ses} offline.
\end{rem} 

\RevTen{One important input for Algorithm \ref{algo:admisibility} is the matrix $M_0$ which acts as an hyper-parameter for the overall system. Note that admissibility of the set of schedules $\Gamma$ only depends on some arbitrary choice of $M_0$ as long as it is positive definite. Hence, setting $M_0=I$ can be considered a systematic default option. On the other hand, the choice of $M_0$ may affect the time until admissibility is concluded in Algorithm \ref{algo:admisibility}. However, optimizing $M_0$ to reduce such time is considered out of the scope of this work but could be explored in future research.}

\begin{algorithm}
 \caption{\texttt{Admissibility}}
 \begin{algorithmic}[1]
 \renewcommand{\algorithmicrequire}{\textbf{Input:}}
 \renewcommand{\algorithmicensure}{\textbf{Output:}}
 \REQUIRE $n,M_0,\{M_\seq\}_{\seq\in\ses},\ses$.\\
 \ENSURE  $\left(\mathbb{S}_0\subset \interior(\mathbb{S}^\sstar_\Gamma\right))$?\\
  \STATE Set $\mathcal{R}\leftarrow\varnothing$.
  \FOR {$\ses'\in\mathcal{P}(\ses)$}
  \IF{$|\ses'|= n$}
  \STATE $\mathcal{R}\leftarrow\mathcal{R}\cup\{$\\ \texttt{IsolatedSolutions}$(n,M_0,\{M_\seq\}_{\seq\in\ses},\ses,\ses')\}$
  \ELSE
  \IF{$|\ses'|< n$}
  \STATE $\mathcal{R}\leftarrow\mathcal{R}\cup\{$ \\\texttt{RegularSolutions}$(n,M_0,\{M_\seq\}_{\seq\in\ses},\ses,\ses')\}$
  \ENDIF
  \ENDIF
  \ENDFOR\\
  \RETURN $\min\mathcal{R}>1$
 
 \end{algorithmic}
 \label{algo:admisibility}
 \end{algorithm}

\begin{algorithm}
 \caption{\texttt{IsolatedSolutions}}
 \begin{algorithmic}[1]
 \renewcommand{\algorithmicrequire}{\textbf{Input:}}
 \renewcommand{\algorithmicensure}{\textbf{Output:}}
 \REQUIRE $n,M_0,\{M_\seq\}_{\seq\in\ses},\ses,\ses'$.\\
 \ENSURE  $r$.\\

  \STATE $\mfb{X}_{\ses'}\leftarrow$ real solutions of $\{\mfb{x}^TM_\seq\mfb{x}=1,\forall \seq\in\ses'\}$ as in Remark \ref{rem:homotopy}.
  \FOR{$\mfb{x}\in\mfb{X}_{\ses'}$}
    \IF{
    exists $\seq\in\ses\setminus\ses'$ such that $\mfb{x}^TM_\seq\mfb{x}<1$}
    \STATE $\mfb{X}_{\ses'}\leftarrow\mfb{X}_{\ses'}\setminus\{\mfb{x}\}$.\\
    \ENDIF
  \ENDFOR
 \RETURN $r\leftarrow\min\{\mfb{x}^TM_0\mfb{x}:\mfb{x}\in\mfb{X}_{\ses'}\}$
 \end{algorithmic}
 \label{algo:measurezero}
 \end{algorithm}

\begin{algorithm}
 \caption{\texttt{RegularSolutions}}
 \begin{algorithmic}[1]
 \renewcommand{\algorithmicrequire}{\textbf{Input:}}
 \renewcommand{\algorithmicensure}{\textbf{Output:}}
 \REQUIRE $n,M_0,\{M_\seq\}_{\seq\in\ses},\ses,\ses'$.\\
 \ENSURE  $r$.\\
  \STATE Set $Y\leftarrow$ real solutions of $\eqref{eq:poly_eq}$ as in Remark \ref{rem:tracedet}\\
  \STATE $\mathcal{R}\leftarrow\varnothing$
  \FOR{$\lambda\in Y$}
    \STATE $G(\lambda)\leftarrow M_0+\sum_{\seq\in\ses'}\lambda_\seq M_\seq$.\\
    \STATE $\mf{g}\leftarrow$ any element of $\textsf{ker}( G(\lambda))$.\\
    \STATE $\mfb{x}\leftarrow\mf{g} \sqrt{\frac{-\sum_{\seq\in\ses'}\lambda_\seq}{{\mf{g}}^TM_0{\mf{g}}}}$.\\
    \IF{ $\mfb{x}^TM_\seq\mfb{x}> 1, \forall\seq\in\ses\setminus\ses'$}
    \STATE $\mathcal{R}\leftarrow\mathcal{R}\cup\{-\sum_{\seq\in\ses'}\lambda_\seq\}$
    \ENDIF
  \ENDFOR
 \RETURN $r\leftarrow\min\mathcal{R}$.
 \end{algorithmic}
 \label{algo:regular}
 \end{algorithm}

\section{Towards optimal scheduling}
\label{sec:balance}

Up to this point, the analysis has been devoted to the stability aspect of Problem \ref{prob:main}. In this section, we focus on the optimality aspect of the problem based on the stability results obtained so far. First, we study the inherent complexity of Problem \ref{prob:main}. Due to the possible conflicting objectives in \eqref{eq:cost_functional}, the most general setting of Problem \ref{prob:main} is reduced to a combinatorial search between the possible scheduling decisions. This makes it hard to expect that exact and efficient solutions for Problem \ref{prob:main} exist except for very particular cases. This issue is formalized in the following result.  
\begin{prop}
\label{th:np}
Problem \ref{prob:main} is {NP}-hard.
\end{prop}

Henceforth, our approach from this point will be to obtain approximate solutions to Problem \ref{prob:main} and study their performance with respect to the optimal solution. Consider $m>0$ sets of schedules $\Gamma^1,\dots,\Gamma^m$ constructed using Algorithm \ref{algo:ses}, leading to $m$ scheduling policies. Theorem \ref{th:stability} ensures that it is possible to choose between any of these scheduling policies without compromising {asymptotic} stability. Hence, instead of deciding between different perception configurations, a sub-optimal schedule for Problem \ref{prob:main} will be constructed by an appropriate decision sequence between the $m$ scheduling policies induced by $\Gamma^1,\dots,\Gamma^m$. 

Using the dynamic programming framework \cite{bertsekas2000}, Algorithm \ref{algo:dynprog} aims to obtain an optimal schedule for Problem \ref{prob:main} when the schedule is constrained to be constructed using the scheduling policies induced by $\Gamma^1,\dots,\Gamma^m$. Note that this algorithm proceeds to evaluate all scheduling policies for the current state $\mf{\bar{x}}_0$ as in line 3. Then, proceeds to compute $\mf{\bar{x}}(t),P(t)$ in line 7 for $t\in[\tau,\min(\tau^+,T_f)]$ which is the interval in which the schedule piece $\gamma$ is active before $t=T_f$, as well as the cost for such interval in line 8. Note that $\mf{\bar{x}}(t),P(t)$ and the cost can be computed explicitly as a result of Proposition \ref{prop:distro} in Appendix \ref{ap:evol}. Then, $\tau^+<T_f$ it means that $\gamma$ is not sufficiently long to fill the entire window $[0,T_f]$. Hence, the remaining cost to go $J^+$ and schedule $p^+$ optimal for $t\in[\tau^+,T_f]$ is obtained by a recursive call in line 10. Otherwise, the terminal cost is added in line 13. Once all $m$ options have been evaluated, $\texttt{dynprog}$ returns the one with the best cost as well as the optimal schedule as in line 16, where $\oplus$ means concatenation of schedules. The optimal properties of this algorithm are detailed in the following. 
\begin{algorithm}
 \caption{\texttt{dynprog}}
 \begin{algorithmic}[1]
 \renewcommand{\algorithmicrequire}{\textbf{Input:}}
 \renewcommand{\algorithmicensure}{\textbf{Output:}}
 \REQUIRE $T_f,\tau,\bar{\mf{x}}_0,P_0$.\\
 \ENSURE $p^*$, $J^*$, $\Gamma^*$.\\
  \STATE $J^*\leftarrow \infty$.
  \STATE $\Gamma^*\leftarrow \varnothing$
  \FOR { $\Gamma$ in $\{\Gamma^1,\dots,\Gamma^m\}$}
\STATE $\seq\leftarrow\seq^\sstar(\mfb{x}_0;\Gamma)$ from \eqref{eq:sw_law}.
  \STATE $\tau^+ \leftarrow \tau+\sum_{k=0}^{\len(\gamma)-1}\Delta^{\gamma_k}$
  \STATE $\alpha\leftarrow \textsf{att}(\gamma,[\tau,\min(\tau^+,T_f)])$

  \STATE Compute $\mf{\bar{x}}(t),P(t)$ for schedule $\gamma$ using $\mf{\bar{x}}(\tau)=\bar{\mf{x}}_0, P(\tau)=P_0$ on \eqref{eq:distro} for $t\in[\tau,\min(\tau^+,T_f)]$\\
  \STATE $J \leftarrow  \frac{\lambda_{\mf x}}{T_f}\displaystyle\int_{\tau}^{\min(\tau^+,T_f)} \mf{\bar{x}}(t)^TQ\mf{\bar{x}}(t)\ +\tr(QP(t)) \nd t$ $+ \frac{\lambda_r}{T_f} \displaystyle\sum_{k=0}^{\alpha-1} r^{\gamma_k}$
  \IF {($\tau^+<T_f$)}
  \STATE $\{p^+,J^+,\Gamma^+\}\leftarrow$\\$\texttt{dynprog}(T_f,\tau^+,\mf{\bar{x}}(\tau^+),P(\tau^+))$
  \STATE $J\leftarrow J+J^+$ 
  \ELSE
  \STATE $J\leftarrow J+{\lambda_{\mf x}}\left(\bar{\mf{x}}(T_f)^TQ_f\bar{\mf{x}}(T_f) + \tr(QP(T_f))\right)$
  \STATE $p^+\leftarrow \varnothing$
  \ENDIF
  \IF{ $J\leq J^*$ }
        \STATE $J^*\leftarrow J$
        \STATE $p^*\leftarrow \gamma\oplus p^+$
        \STATE $\Gamma^*\leftarrow \Gamma$
  \ENDIF
  \ENDFOR
 
 \end{algorithmic}
 \label{algo:dynprog}
 \end{algorithm}
\begin{prop}
\label{prop:dynprog}
Calling $\texttt{\emph{dynprog}}(T_f,0,\mf{\bar{x}}_0,P_0)$ obtains the optimal schedule and cost $p^*,J^*$ for Problem \ref{prob:main} for the case when the schedule is constrained to be constructed using the scheduling policies for $\{\Gamma^1,\dots,\Gamma^m\}$. Moreover, the worst case complexity of Algorithm \ref{algo:dynprog} is $O\left(m^{\lfloor T_f/c\rfloor}\right)$ where $c=\min\left\{\sum_{k=0}^{\len(\gamma)-1}\Delta^{\gamma_k}:\gamma \in\Gamma^1\cup\cdots\cup\Gamma^m\right\}$.
\end{prop}
\begin{rem}
\label{rem:np}
Due to NP-hardness of Problem \ref{prob:main}, it is not surprising that the dynamic programming approach for this problem leads to an algorithm that does not have a polynomial complexity in $T_f$. However, these types of algorithms, in which unrolling the recursive calls lead to an exponentially growing tree as in Algorithm \ref{algo:dynprog}, are widely studied. Hence, some performance improvements in terms of complexity and run-time reduction can be made to Algorithm \ref{algo:dynprog} such as applying a branch-and-bound technique \cite[Chapter 2.3.3]{bertsekas2000}. In addition, heuristics and approximate solutions such as limited look-ahead policies, roll-out algorithms, among others \cite[Chapter 6]{bertsekas2000} can be applied in practice. For instance, a balanced $\sptwo$ strategy can be applied where the set of schedules $\Gamma$ in line 8 of Algorithm \ref{algo:sp2} is selected by calling  $\{-,-,\Gamma\}=\texttt{dynprog}(T,0,\hat{\mf{x}}[k|k-1],\hat{P}[k])$ as a moving horizon strategy with a look-ahead window of size $T$.
\end{rem}

Nevertheless, all heuristics and approximate approaches mentioned before will have a relative performance loss with respect to the optimal schedule for Problem \ref{prob:main} only when the solution is constrained to be constructed using the scheduling policies induced by $\Gamma^1,\dots,\Gamma^m$, and not necessarily the real optimum of Problem \ref{prob:main}. Despite this, in the following result, we show that the solution of Algorithm \ref{algo:dynprog} can approximate the optimal performance for sufficiently large number of sets of schedules $m$, increasing the computational power applied.

\begin{thm}
\label{th:optimal}
Let $p^*$ be the optimal cost for the general setting of Problem \ref{prob:main} and assume that \eqref{eq:switch} is stabilizable. Then, for any compact sets $B_{\mf{\bar{x}}}\subset\mathbb{R}^n,B_{P}\subset\mathbb{R}^{n\times n}$ and any $\varepsilon>0$ there exists $\Gamma^1,\dots,\Gamma^m$ with $m<\infty$ such that 
$$
|\mathcal{J}(p^*;\mf{\bar{x}}_0,P_0)-\mathcal{J}(p;\mf{\bar{x}}_0,P_0)|<\varepsilon
$$ 
with $p$ from by Algorithm \ref{algo:dynprog}, for any $\mf{\bar{x}}_0\in B_\mf{\bar{x}}$, $P_0\in B_P$.
\end{thm}

\section{Simulation examples}
\label{sec:examples}

\subsection{Double integrator}
\label{sec:doubleIntegrator}
\begin{figure}
    \centering
    \includegraphics[width=0.4\textwidth]{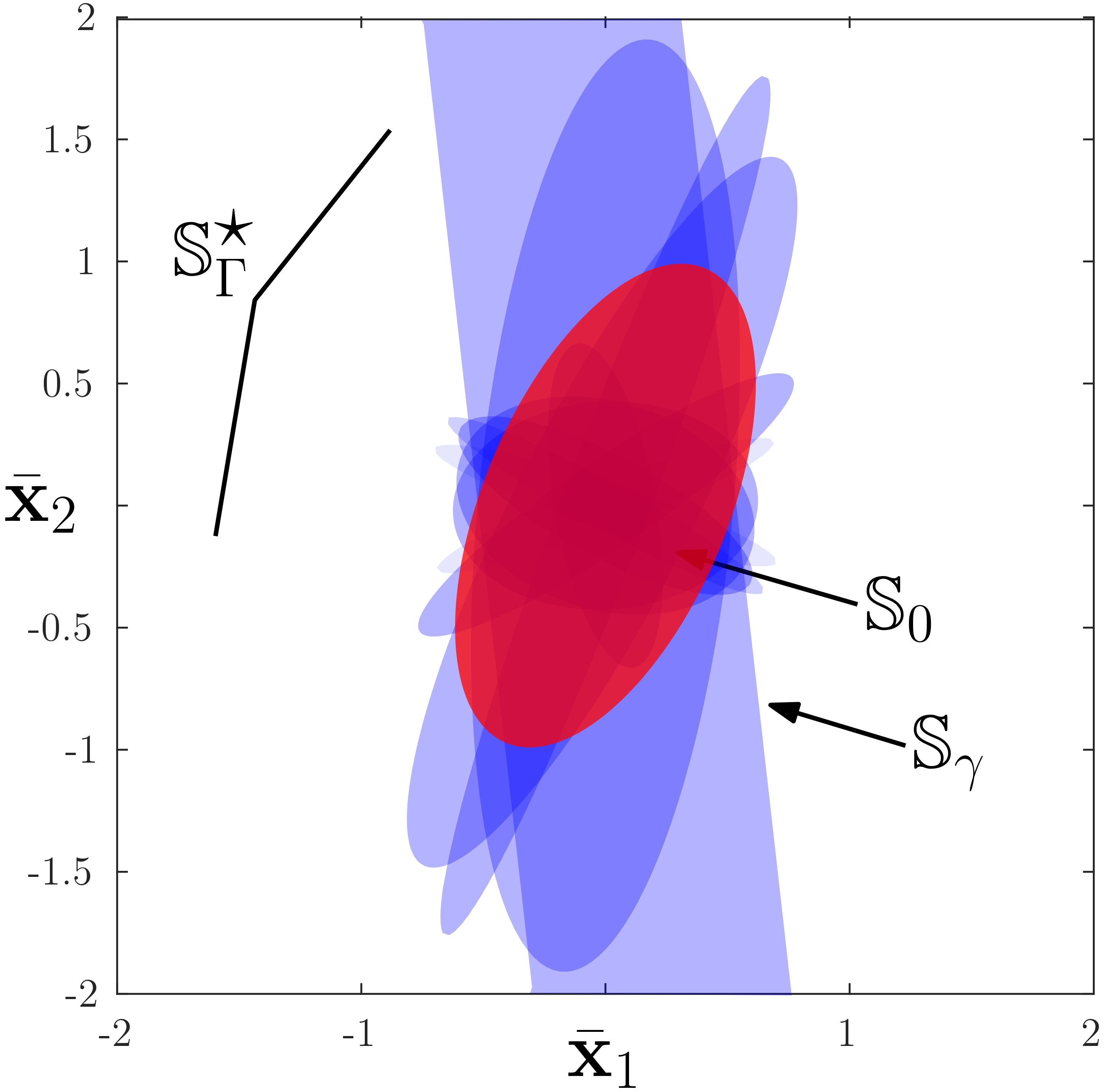}
    \caption{Set $\cset_{\ses}^\sstar$ for the set of schedules $\ses$ described in Section \ref{ex:double} as well as $\cset_0$ to show admissibility.} 
    \label{fig:admisibility}
\end{figure}
\begin{figure}[htp]
    \centering
    \includegraphics[width=0.4\textwidth]{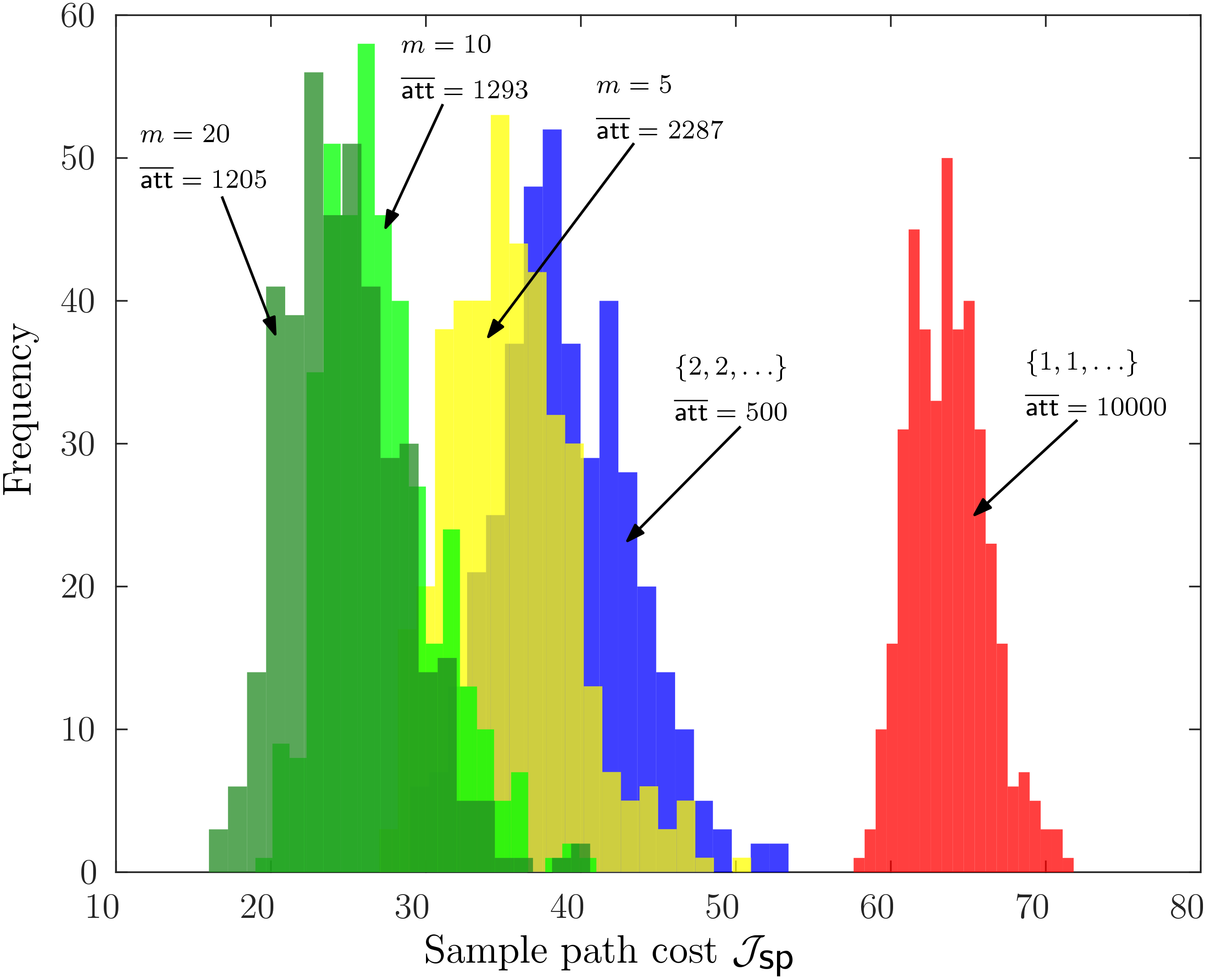}
    \caption{Resulting histograms for the sample path cost $\mathcal{J}_{\mathsf{sp}}$ for example of Section \ref{ex:double} using the balanced $\sptwo$ strategy with $m=5,m=10,m=20$ and static schedules $\{1,1,\dots\}, \{2,2,\dots\}$.} 
    \label{fig:trajs2d}
\end{figure}
\label{ex:double}
As a first example consider system \eqref{eq:system} with
$$
A=\begin{bmatrix}
0 & 1 \\
0 & 0
\end{bmatrix}, \ B=\begin{bmatrix}
0\\
1
\end{bmatrix}, \ C = [1,0],\  W_0 = 1
$$ 
as well as $D=2$ with perception latencies $\Delta^1=0.01, \Delta^2=0.1$ with their corresponding covariances $\Sigma^1=0.5, \Sigma^2=0.01$. In addition, consider
$$
L^1=L^2= [-1.5   -3], 
M_0 = \begin{bmatrix}
    3.53  & -1.10\\
   -1.10  &  1.36
\end{bmatrix}
$$
\RevSeven{where $M_0$ was chosen as an arbitrarily positive definite matrix.} To depict admissibility, Algorithm \ref{algo:ses} was used with $\ell=20$ to generate a set of schedules $\Gamma$ and Algorithm \ref{algo:admisibility} was used to check admissibility. Note that in Algorithm \ref{algo:admisibility} the only meaningful subsets of schedules $\ses'$ in the power set $\mathcal{P}(\ses)$ are $\ses'$ containing either a single schedule, or pairs of schedules as a result of the discussion in Remark \ref{rem:homotopy}. Moreover, note that in the case of $\ses'$ containing a single schedule $\seq$, \eqref{eq:Gmat} results in $G(\lambda)=M_0+\lambda M_\seq$ and therefore possible solutions of Lagrange multipliers $\lambda\in\mathbb{R}$ are eigenvalues of $M_0^{-1}M_\seq$. The resulting nullity of $G(\lambda)$ for any $\ses'$ in this experiment was 1 and hence regular critical points were computed as in \eqref{eq:null_space}. In this case, there are no non-regular points, since for $|\ses'|=1$ there is a single vector $\nabla h_\seq(\mfb{x})$. For the case of $|\ses'|=n=2$ critical points were obtained by solving a polynomial system of equations given by the intersection of the two resulting ellipses as in Remark \ref{rem:homotopy}. Hence, in this setting the admissibility value was obtained to be $R=1.142$ which indicates that $\ses$ is admissible. Figure \ref{fig:admisibility} shows $\cset_{\ses}^\sstar$ and $\cset_0$ for this example where it shown that $\cset_0\subset\interior(\cset_{\ses}^\sstar)$. Moreover, note that none of the individual sets $\mathbb{S}_\gamma, \gamma\in\Gamma$ manage to cover $\mathbb{S}_0$ by separate.

Now, we show the performance of an heuristics implementation of the strategy presented in Section \ref{sec:balance}. To do so, we simulated system \eqref{eq:system} using explicit Euler-Maruyama method with time step $h=10^{-5}$ over the interval $[0,T_f=100]$ for each experiment. A balanced $\sptwo$ strategy as described in Remark \ref{rem:np} was used with limited look-ahead window of size $T = 2$. The performance is evaluated using the cost \eqref{eq:cost_functional} for each sample path with $Q=Q_f=\diag([2,1]), \lambda_{\mf{x}}=1$, $\lambda_r=0.05$. Moreover, $r^1=r^2=1$ such that the number of sampling events is penalized in the cost. Furthermore, 400 sample paths where generated with $\mf{\bar{x}}_0=[1,1], P_0 = I$ for $m=5,m=10,m=20$ respectively.   For comparison, the same amount of sample paths where obtained for schedules $\{1,1,\dots\},\{2,2,\dots\}$ respectively to study the performance of the classical approach of maintaining the same perception configuration during the whole experiment. The resulting histograms for the sample path cost $\mathcal{J}_{\textsf{sp}}$ for the whole interval $[0,T_f]$ are shown in Figure \ref{fig:trajs2d} for each case. It can be observed that the average cost of our approach is reduced when compared to the static approaches. Moreover, the results suggest that increasing the number of sets of schedules $m$ improves the  cost $\mathcal{J}_{\textsf{sp}}$ and  average attention $\overline{\textsf{att}}$. 

\subsection{Particle mobile robot}
\label{ex:particle}
Now consider a system with state $\mf{x}=[\mf{x}_1,\dots,\mf{x}_6]^T$, and dynamics given $\dot{\mf{x}}_1=\mf{x}_2, \dot{\mf{x}}_3=\mf{x}_4, \dot{\mf{x}}_5=\mf{x}_6$ as well as $\dot{\mf{x}}_2 = \mf{u}_1/\mu + \mf{w}_1, \dot{\mf{x}}_4 = \mf{u}_2/\mu+ \mf{w}_2, \dot{\mf{x}}_6 = \mf{u}_3/\mu+ \mf{w}_3$ with $\mu=0.1$. This system corresponds to the model of a particle mobile robot with mass $\mu$. This kind of system has been useful as a model for aerial vehicles such as multi-rotors in \cite{mellinger,Faessler18ral,aldana2021} were $[\mf{x}_1,\mf{x}_3,\mf{x}_5]^T$ is the position of the multi-rotor and $[\mf{x}_2,\mf{x}_4,\mf{x}_6]^T$ is the velocity. In fact, a common strategy for the multi-rotor control is adopt a hierarchical control approach, where control signals $\mf{u}_1,\mf{u}_2,\mf{u}_3$ are designed exclusively for position as a high-level controller. In a second stage, the resulting position controls are used to construct a low level controller for the attitude of the multi-rotor. Moreover, measurements for position are taken as $C\mf{x} = [\mf{x}_1,\mf{x}_3,\mf{x}_5]^T$ using visual perception \cite{aldana2021}. In this example we adopt the noise model from \cite[Equation (9)]{luca2020} such that given a perception latency $\Delta$ the resulting measurement covariance is  $\Sigma(\Delta) = (b/\Delta)I$ with $b=0.2$. {This model is reasonable for demonstrative purposes in this work. However, in a more practical setting,  $\Sigma(\Delta)$ must be characterized according to the relation between latency and precision of the actual perception algorithms and sensors used, following a similar procedure to the one described in \cite{tcst_codesign}.} In this case, even when the measurement covariance is reduced as $\Delta$ is increased, the disturbance covariance $W_d(\Delta)$, given in \eqref{eq:covarianceW} from Appendix \ref{ap:evol}, increases.

\begin{figure}
    \centering
    \includegraphics[width=0.4\textwidth]{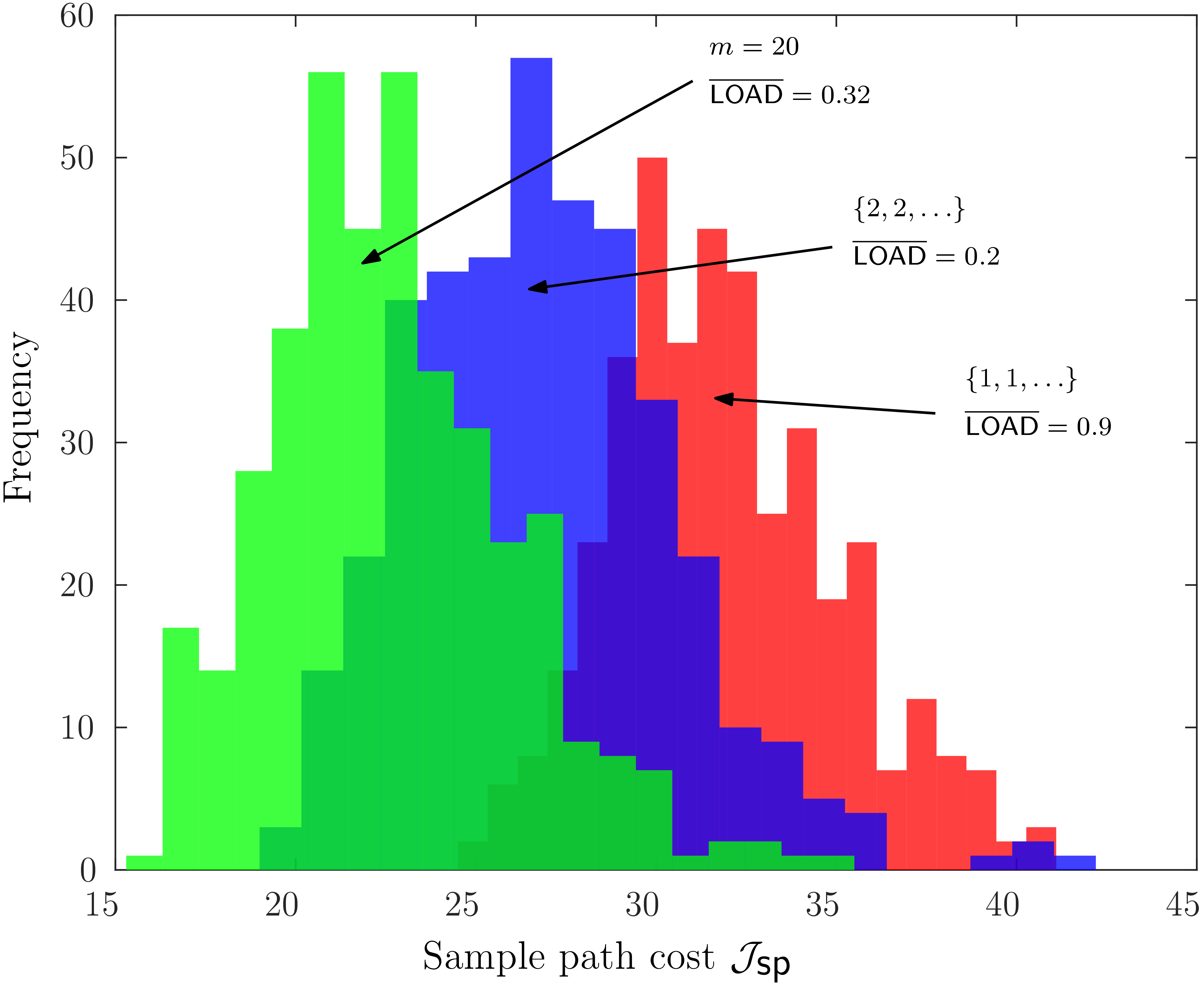}
    \caption{Resulting histograms for the sample path cost $\mathcal{J}_{\mathsf{sp}}$ for the example of Section \ref{ex:double} using the balanced $\sptwo$ strategy with $m=20$ and static schedules $\{1,1,\dots\}, \{2,2,\dots\}$.} 
    \label{fig:trajs3d}
\end{figure}

Now, we test our scheduling approach assuming only two different latencies $\Delta^1=1/30,\Delta^2=4/30$ corresponding to the frame-rate of a typical camera or 4 times the frame-rate. Let $\Sigma^1=\Sigma(\Delta^1), \Sigma^2=\Sigma(\Delta^2)$ with the covariance model $\Sigma(\Delta)$ described before and $W_0 = 0.5I$. Moreover, consider penalties $r^1 = 0.9\Delta^1, r^2 = 0.2\Delta^2$ which correspond to CPU loads of $90\%$ and $20\%$ respectively for each method. In addition, the parameters $Q=Q_f=\diag([2,1,2,1,2,1]), \lambda_{\mf{x}}=1$, $\lambda_r=0.1$ were chosen. The $\sptwo$ strategy was tested similarly as in the example of Section \ref{ex:double} with time step $h=10^{-5}$, $T_f=100$ and limited look-ahead window length $T = 10$. Figure \ref{fig:trajs3d} shows the resulting histograms for sample paths with $\mf{\bar{x}}_0=[1,1,1,1,1,1], P_0=I$ for our approach with $m=20$ and static schedules $\{1,1,\dots\}, \{2,2,\dots\}$ with $400$ sample paths in each case. It is observed that our approach outperforms the static approaches in the sample cost $\mathcal{J}_{\textsf{sp}}$. Moreover, the average CPU load $\overline{\textsf{LOAD}}$ is reduced with respect to the worst case value of $90\%$.

\section{Conclusion}
This work studied a perception scheduling approach to tackle the latency-precision trade-off, where measurements can be taken using several perception mode possibilities with different noise levels. It was shown that the stability of the closed-loop model is tied to the stability of a switching system. Moreover, the latency-precision trade-off was modeled by means of a cost function taking into account control precision and a penalty for each perception mode to incorporate energetic and CPU load points of view. Three main tools were proposed in this context to solve the problem: scheduling policy candidate construction using switching systems theory, an admissibility checking algorithm for policy candidates, and sub-optimal strategies based upon the scheduling policy candidates. \RevOne{Different from general switching systems theory, we take advantage of the particular structure of the problem to provide new insights on the construction of multiple scheduling policies for optimization purposes and on its theoretical complexity.} Finally, some illustrative examples were provided, including one application for a mobile robot. \RevOne{These results motivate future research on this problem with other noise models for the perception method or imperfect timing in the perception latency.}
\appendix

\section{Distribution, estimation and cost of $\mf{x}(t)$}
\label{ap:evol}

In this section, we obtain expressions for $\hat{\mf{x}}[k|k-1]:=\mathbb{E}\{\mf{x}[k]|\mf{z}[0],\dots,\mf{z}[k-1]\}$, the distribution of $\mf{x}(t)$ as the closed-loop solution of \eqref{eq:system} and an expression for \eqref{eq:cost_functional}.

\begin{prop}
\label{prop:estimator}
Consider Assumptions \ref{as:dist} and \ref{as:meas}. Then, given measurements $\mf{z}[0],\dots,\mf{z}[k-1]$ and a perception schedule $p$, the value of $\hat{\mf{x}}[k|k-1]:=\mathbb{E}\left\{\mf{x}[k]\ \big|\ \mf{z}[0],\dots,\mf{z}[k-1]\right\}$ is computed according to:
\begin{equation}
    \begin{aligned}
    H[k] &= A_d(\Delta^{p_k})\hat{P}[k]C^T\left(C\hat{P}[k]C^T+\Sigma^{p_k}\right)^{-1}\\
    \hat{\mf{x}}[k+1|k] &= A_d(\Delta^{p_k})\hat{\mf{x}}[k|k-1]+B_d(\Delta^{p_k})\mf{u}[k] \\&
    + H[k](\mf{z}[k]-C\hat{\mf{x}}[k|k-1])\\
    \hat{P}[k+1] &= (A_d(\Delta^{p_k})-H[k]C)\hat{P}[k]A_d(\Delta^{p_k})^T\\&+W_d(\Delta^{p_k}) 
    \end{aligned}
\end{equation}
with $\hat{\mf{x}}[0|-1]=\bar{\mf{x}}_0, \hat{P}[0]=P_0$,  $A_d(\tau):=\exp(A\tau), B_d(\tau):=\int_0^\tau A_d(s)\nd s B, W_d(\tau):=\int_0^\tau A_d(s)W_0A_d(s)^T\nd s$. Moreover, $\cov\{\hat{\mf{x}}[k|k-1]-\mf{x}[k]\}=\hat{P}[k]$.
\end{prop}

\begin{pf}
First, note that the explicit solution to \eqref{eq:system} in the interval $t-\tau_k\in[0,\Delta^{p_k})$ is given by
\begin{equation}
\label{eq:discrete_model}
    \mf{x}(t) = A_d(t-\tau_k)\mf{x}[k] + B_d(t-\tau_k)\mf{u}[k] + \mf{w}_d(t)
\end{equation}
with $\mf{w}_d(t)\sim\mathcal{N}(0,W_d(t-\tau_k)$ \cite[Section 4.5.2]{Soderstrom}. Note that evaluating \eqref{eq:discrete_model} at $t=\tau_{k+1}$ leads to the discrete-time system which transitions from $\mf{x}[k]$ to $\mf{x}[k+1]$. Thus, the result follows by applying the estimator in \cite[Theorem 4.1, Page 228]{astrom}.
\end{pf}
\begin{prop}
\label{prop:distro}
Let $p$ a perception schedule and $\Lambda(t-\tau_k) := A_d(t-\tau_k) + B_d(t-\tau_k)L^{p_k}, t-\tau_k\in[0,\Delta^{p_k})$. Moreover, consider Assumptions \ref{as:dist} and \ref{as:meas}. Then, the solution $\mf{x}(t)$ of the SDE in \eqref{eq:system} under the controller $\mf{u}[k]=L^{p_k}\mf{\hat{x}}[k|k-1]$ given samples $\{\mf{z}[0],\dots,\mf{z}[k-1]\}$ is
\begin{equation}
\label{eq:discrete_model_aut}
\begin{aligned}
    \mf{x}(t) &= \Lambda(t-\tau_k)\mf{x}[k] +\mf{w}_\Lambda(t)
\end{aligned}
\end{equation}
for $t-\tau_k\in[0,\Delta^{p_k})$ and $\mf{w}_\Lambda(t)\sim\mathcal{N}(0,W_\Lambda(t-\tau_k))$ with:
\begin{equation}
\label{eq:covarianceW}
\begin{aligned}
W_\Lambda(t-\tau_k) &= B_d(t-\tau_k)L^{p_k}\hat{P}[k]{L^{p_k}}^T B_d(t-\tau_k)^T\\&+W_d(t-\tau_k)
\end{aligned}
\end{equation}
Moreover, $\mf{x}(t)\sim\mathcal{N}(\mf{\bar{x}}(t),P(t))$ for $t-\tau_k\in[0,\Delta^{p_k})$ with 
\begin{equation}
\label{eq:distro}
\begin{aligned}
\mf{\bar{x}}(t) &= \Lambda(t-t_k)\mf{x}[k], \ \  \mf{\bar{x}}[0] = \mf{\bar{x}}_0 \\
P(t) &= \Lambda(t-\tau_k)P[k]\Lambda(t-\tau_k)^T+W_\Lambda(t-\tau_k)
\end{aligned}
\end{equation}
with $P[0] = P_0$. In addition, with $\alpha:=\textsf{att}(p;[0,T_f])$, cost \eqref{eq:cost_functional} is
\begin{equation}
\begin{aligned}
    \label{eq:cost_functional_det}
    &\mathcal{J}(p;\mf{\bar{x}}_0,P_0) =  \frac{\lambda_{\mf x}}{T_f}\int_{0}^{T_f} \mf{\bar{x}}(t)^TQ\mf{\bar{x}}(t)\ +\tr(QP(t)) \nd t  \\
    & +{\lambda_{\mf x}}\left(\bar{\mf{x}}(T_f)^TQ_f\bar{\mf{x}}(T_f) + \tr(Q_fP(T_f))\right)+ \frac{\lambda_r}{T_f} \sum_{k=0}^{\alpha-1} r^{p_k}
    \end{aligned}
 \end{equation}
\end{prop}

\begin{pf}
Proposition \ref{prop:estimator} implies that $\hat{\mf{x}}[k|k-1]=\mf{x}[k]+\tilde{\mf{x}}[k]$ where $\tilde{\mf{x}}[k]\sim\mathcal{N}(0,\hat{P}[k])$ given samples $\{\mf{z}[0],\dots,\mf{z}[k-1]\}$. Hence, using $\mf{u}[k]=L^{p_k}\mf{x}[k] + L^{p_k}\mf{\tilde{x}}[k]$ in \eqref{eq:discrete_model} leads directly to \eqref{eq:discrete_model_aut} with $\mf{w}_\Lambda=B_d(t-\tau_k)L^{p_k}\tilde{\mf{x}}[k] + \mf{w}_d(t)$. Moreover, according to \cite[Section 4.5.2]{Soderstrom} the process $\{\mf{w}_d[0],\dots,\mf{w}_d[k],\mf{w}_d(t)\}$ is a white noise process and thus, its individual random variables are uncorrelated. Therefore, $\tilde{\mf{x}}[k]$ and $\mf{w}_d(t)$ are uncorrelated. Henceforth, computing the covariance of $\mf{w}_\Lambda(t)$ leads to \eqref{eq:covarianceW}. Furtheremore, direct computation of the expectation and covariance over \eqref{eq:discrete_model_aut} leads to \eqref{eq:distro}. Finally, \eqref{eq:cost_functional_det} is computed as in \cite[Theorem 1]{aldana2020}.
\end{pf}

\section{$C^\sstar$ sets and gauge functions}
\label{ap:cstar}
\begin{defn}\label{def:cstar}\cite[Definition 1]{jungers2014} A $C^\sstar$ set $\cset\subset\mathbb{R}^n$ is a compact, star-convex set with the origin as a center i.e., for any $\mf{x}\in\cset$ the whole line 
$\{\mf{x}_\lambda\in\mathbb{R}^n: \mf{x}_\lambda = \lambda \mf x \andset \lambda\in[0,1] \}
$
is contained in $\cset$ and $0\in\interior(\cset)$. Moreover, the gauge function of a $C^\sstar$ set $\cset$ is defined as
\begin{equation}
    \Psi(\mf{x};\cset) := \inf\{\alpha\in\mathbb{R}: \alpha\geq 0 \andset \mf{x}\in\alpha \cset\}
\end{equation}
\end{defn}
This is, $\Psi(\mf{x};\cset)$ is the smallest scale $\alpha$ such that $\mf{x}$ is still contained in $\alpha \cset$. In the following, we enumerate some properties regarding $C^\sstar$ sets and their gauge functions:
\begin{lem}
\label{le:cstar}
Let $\cset$ be a $C^\sstar$ set and $\Psi(\bullet;\cset):\mathbb{R}^n\to[0,+\infty)$ be its gauge function. Then, the following statements are true:
\begin{enumerate}[1)]
    \item Let $0<\alpha<1$ and $\mf{x}\in \cset$, then $\alpha\mf{x}\in \cset$.
    \item Let $0<\alpha<\beta$, then $\alpha \cset\subset \beta \cset$.
    \item Let $\alpha:=\Psi(\mf{x};\cset)$, then $\mf{x}\in\partial(\alpha \cset)$.
    \item $\Psi(\bullet;\cset)$ is homogeneous of degree one, i.e. $\Psi(\beta \mf{x};\cset) = \beta \Psi(\mf{x};\cset),\ \forall\beta\geq 0$ and $\forall\mf{x}\in\mathbb{R}^n$.
    \item Let $\cset'$ be a $C^\sstar$ set with $\cset'\subset \cset$. Thus $\Psi(\mf{x};\cset')>\Psi(\mf{x};\cset), \forall \mf{x}\in\mathbb{R}^n$.
    \item Let $\cset_0=\{\mf{x}\in\mathbb{R}^n:\mf{x}^TM_0\mf{x}\leq 1\}$ for some positive definite matrix $M_0$. Then, $\Psi(\mf{x};\cset_0) = \sqrt{\mf{x}^TM_0\mf{x}}$ for any $\mf{x}\in\mathbb{R}^n$.
\end{enumerate}
\end{lem}

\begin{pf}
These properties can be verified equivalently as the properties of gauge functions of standard convex sets \cite[Page 28]{convex}.
\end{pf}

\section{Proofs}
\label{sec:proofs}

\subsection{Proof of Theorem \ref{th:stability}}
\label{ap:stability}

In order to show Theorem \ref{th:stability} we first provide 2 auxiliary technical lemmas:
\begin{lem}
\label{le:invariant_set}
Let $\ses$ be an admissible set of schedules and let $\seq^\sstar:=\seq^\sstar(\mfb{x}_0;\ses)$ be a schedule obtained from \eqref{eq:sw_law} for the initial condition $\mfb{x}_0\in\cset_\ses^\sstar$. Thus, if the $\sptwo$ strategy is used in \eqref{eq:switch} then $\mfb{x}[\len(\seq^\sstar)]\in\cset_0\subseteq\text{\normalfont int}\left(\cset_\ses^\sstar\right)$.
\end{lem}

\begin{pf}
Since $\mfb{x}_0\in\cset_\ses^\sstar$ then there exist a non empty set $\ses'\subseteq \ses$ such that $\mfb{x}_0\in\cset_\seq\ $  $\forall \seq\in{\ses}'$. Moreover, $\mfb{x}_0^TM_\seq\mfb{x}_0\leq 1$ for any $\seq\in\ses'$ and $\mfb{x}_0^TM_\seq\mfb{x}_0>1$ for $\seq\in \ses\setminus \ses'$. Hence, $\seq^\sstar=\seq^\sstar(\mfb{x}_0;\ses)\in \ses'$ and $\mfb{x}_0\in\cset_{\seq^\sstar}$. Therefore, by construction of the set $\cset_{\seq^\sstar}$, system \eqref{eq:switch} after $\len(\seq^\sstar)$ steps  will result in $\mfb{x}[\len(\seq^\sstar)]\in\cset_0\subset \cset_\ses^\sstar$.
\end{pf}

\begin{lem}
\label{le:lyapunov}
Let $\ses$ be an admissible set of schedules, $\Psi(\bullet;\cset_\ses^\sstar)$ be its corresponding gauge function and let $\seq^\sstar:=\seq^\sstar(\mfb{x}_0;\ses)$ be a schedule obtained from \eqref{eq:sw_law} for the initial condition $\mfb{x}_0\neq 0$. Thus, if the $\sptwo$ strategy is used in \eqref{eq:switch} then
\begin{enumerate}
    \item $
\Psi(\mfb{x}[\len(\seq^\sstar)];\cset_\ses^\sstar) < \Psi(\mfb{x}_0;\cset_\ses^\sstar)
$
\item $
\Psi(\mfb{x}[\len(\seq^\sstar)];\cset_0) < \Psi(\mfb{x}_0;\cset_0)
$
\end{enumerate}
\end{lem}

\begin{pf}
Note that by Lemma \ref{le:cstar}-3) in Appendix \ref{ap:cstar} it is obtained that $\mfb{x}_0\in\partial(\alpha_0 \cset_\ses^\sstar)$ with $\alpha_0:=\Psi(\mfb{x}_0;\cset_\ses^\sstar)$. Similarly, one can obtain that $\mfb{x}[\len(\seq^\sstar)] \in \partial(\alpha\cset_\ses^\sstar)$ with $\alpha:=\Psi(\mfb{x}[\len(\seq^\sstar)];\cset_\ses^\sstar)$. Moreover, since $\mfb{x}_0/\alpha_0\in\partial\cset_\ses^\sstar$, then Lemma \ref{le:invariant_set} implies that $\mfb{x}[\len(\seq^\sstar)]/\alpha_0\in\cset_0\subseteq \interior(\cset_\ses^\sstar)$. Hence, $\partial(\alpha\cset_\ses^\sstar)\subset \interior(\alpha_0\cset_\ses^\sstar)$ strictly, unless $\alpha_0=0$ which is not possible since $\mfb{x}_0\neq 0$. However, for that to be true, we require $\alpha<\alpha_0$ which is exactly item 1). For item 2),  since $\ses$ is admissible then $\cset_0\subset\cset_\ses^\sstar$ and therefore $1=\Psi(\mf{x}_0/\alpha_0;\cset_\ses^\sstar)<\Psi(\mfb{x}_0/\alpha_0;\cset_0)$ by Lemma \ref{le:cstar}-5). Equivalently $\alpha_0<\Psi(\mfb{x}_0;\cset_0)$. Recall that $\mfb{x}[\len(\seq^\sstar)]/\alpha_0\in\cset_0$ and therefore $\Psi(\mfb{x}[\len(\seq^\sstar)]/\alpha_0;\cset_0)\leq1$ or equivalently $\Psi(\mfb{x}[\len(\seq^\sstar)];\cset_0)\leq\alpha_0<\Psi(\mfb{x}_0;\cset_0)$.
\end{pf}

Now, we are in position to prove Theorem \ref{th:stability}. Note that as a consequence of the previous lemma, both $\Psi(\bullet;\cset_\ses^\sstar)$ and $\Psi(\bullet;\cset_0)$ are Lyapunov function candidates for \eqref{eq:switch}.  Note that $\Psi(\bullet;\cset_\ses^\sstar)$ can be used for a similar setting. However, we use $\Psi(\bullet;\cset_0)$ instead to take advantage of the fact that this Lyapunov function is independent of the set of schedules used. Henceforth, switching between admissible sets of schedules as in line 8 of Algorithm \ref{algo:sp2} is possible. Let $\left\{\ell_\kappa\right\}_{\kappa=0}^\infty$ be the sequence of switching time instants obtained by using the strategy $\sptwo$ on \eqref{eq:switch}. Moreover, define $\mf{y}[\kappa]:=\mfb{x}[\ell_\kappa]$ as the state at instants where the schedule changes. Consider the Lyapunov function candidate $V(\mf{y}[\kappa]) := \Psi(\mf{y}[\kappa];\cset_0)$. Lemma \ref{le:lyapunov} implies $V(\mf{y}[\kappa+1])- V(\mf{y}[\kappa])<0$ which in turn implies $\lim_{\kappa\to\infty} \mf{y}[\kappa] = \lim_{\kappa\to\infty} \mfb{x}[\ell_\kappa] = 0$. For any $\mfb{x}[k]$ with $k\notin\{\ell_\kappa\}_{\kappa=0}^\infty$, i.e. values of $\mfb{x}$ not at schedule change instants, it can be obtained from \eqref{eq:switch} that
$
\mfb{x}[k] = \Lambda(\Delta^{p_{k-1}})\Lambda(\Delta^{p_{k-2}})\cdots \Lambda(\Delta^{p_{\ell'}}) \mfb{x}[\ell']
$
with $\ell'=\max\{\ell\in\{\ell_\kappa\}_{\kappa=0}^\infty:  \ell<k\}$. Hence $\|\bar{\mf{x}}[k]\|\leq \rho_k \|\mfb{x}[\ell']\|$ with $\rho_k$ the spectral radius of $\Lambda(\Delta^{p_{k-1}})\Lambda(\Delta^{p_{k-2}})\cdots \Lambda(\Delta^{p_{\ell'}})$, which since $\lim_{\ell'\to\infty}\mfb{x}[\ell']=\lim_{\kappa\to\infty}\mf{y}[\kappa]=0$ implies that $\lim_{k\to\infty}\mfb{x}[k]=0$ too.

\subsection{Proof of Theorem \ref{eq:optim_theo}}

\label{app:non_convex_equiv}
First, note that due to Lemma \ref{le:cstar}-6) in Appendix \ref{ap:cstar}, $R>1$ if and only if $\Psi(\mfb{x};\cset_0)> 1, \forall \mfb{x}\in\partial \cset_\ses^\sstar$, where $\Psi(\bullet;\ast)$ is the gauge function given in Definition \ref{def:cstar} from Appendix \ref{ap:cstar}. Now, we show that if $R>1$ then $\cset_0\subset \cset_\ses^\sstar$. We proceed by contradiction: let's assume that there exists $\mfb{x}\in\cset_0$ such that $\Psi(\mfb{x};\cset_\ses^\sstar)> 1$. Let $\beta:=\Psi(\mfb{x};\cset_\ses^\sstar)$ and thus $\mfb{x}/\beta\in\partial \cset_\ses^\sstar$ due to Lemma \ref{le:cstar}-3).  Henceforth, since $R>1$, then $\Psi(\mfb{x}/\beta;\cset_0)> 1$ or equivalently $\Psi(\mfb{x};\cset_0)>\beta>1$ due to Lemma \ref{le:cstar}-4) and due to the assumption $\beta=\Psi(\mfb{x};\cset_\ses^\sstar)>1$. However, for $\mfb{x}$ to belong in $\cset_0$ we require $\Psi(\mfb{x};\cset_0)<1$ which leads to a contradiction. Therefore, $\beta=\Psi(\mfb{x};\cset_\ses^\sstar)\leq 1$ for all $\mfb{x}\in\cset_0$. Hence, $\mfb{x}\in \partial(\beta\cset_\ses^\sstar)\subseteq\beta\cset_\ses^\sstar\subseteq\cset_\ses^\sstar$ due to Lemma \ref{le:cstar}-3) and Lemma \ref{le:cstar}-1) for any $\mfb{x}\in\cset_0$ and thus $\cset_0\subseteq \cset_\ses^\sstar$. Finally, note that since all points $\mfb{x}\in\partial\cset_\ses^\sstar$ have $\mfb{x}^TM_0\mfb{x}>1$ (by the assumption   $R>1$), then they do not belong to $\cset_0$. Thus, $\cset_0\cap\partial\cset_\ses^\sstar=\varnothing$ and then $\cset_0\subset\interior(\cset_\ses^\sstar)$. Now, we show that if $\cset_0\subset \interior(\cset_\ses^\sstar)$, then $R>1$. From here, note that $\cset_0\subseteq \interior(\cset_\ses^\sstar)$. We proceed by contradiction: assume that there exists $\mfb{x}\in\partial\cset_\ses^\sstar$ such that $\mfb{x}^TM_0\mfb{x}\leq 1$. Hence, $\mfb{x}\in\cset_0\subseteq\interior(\cset_\ses^\sstar)$ which is a contradiction. Therefore, $R>1$ for all $\mfb{x}\in\partial\cset_\ses^\sstar$ which concludes the proof.

\subsection{Proof of Corollary \ref{le:subproblem}}
\label{app:subproblem}

Let $\mfb{x}^*$ be the critical point of \eqref{eq:program} which leads to the global minimum $R$ and note that $\mfb{x}^*\in\partial\cset_\ses^\sstar$. Hence, $\mfb{x}^*\in\partial\cset_\seq$ for at least one $\seq\in\ses$ and at most all $\seq\in\ses$. Let $\ses'$ be the set of all those $\seq\in\ses$ and note that $\ses'\in \mathcal{P}(\ses)$. Therefore, $(\mfb{x}^*)^TM_\seq(\mfb{x}^*)=1$ for all $\seq\in\ses'$ and is a local minimum of \eqref{eq:subproblem} for such $\ses'$. Since $\mfb{x}^*\in\partial\cset_\ses^\sstar$ and $\mfb{x}^*\in\partial\cset_\seq$ only for $\seq\in\ses'$, thus $\mfb{x}^*\notin\cset_\seq$ for $\seq\in\ses\setminus \ses'$, and thus $\mfb{x}^*\in\mfb{X}_{\ses'}$. Now, for $\mfb{x}^*$ to be the global minimum, critical points $\mfb{x}$ comply $(\mfb{x}^*)^TM_0(\mfb{x}^*)\leq\mfb{x}^TM_0\mfb{x}$ in \eqref{eq:subproblem3}.

\subsection{Proof of Theorem \ref{th:regular}}
\label{ap:regular}
In order to show Theorem \ref{th:regular} we proceed by duality analysis, which is summarized in the following two technical lemmas.

\begin{lem}
\label{le:xdag}
Any $\mfb{x}^*$ which is a regular critical point of program \eqref{eq:subproblem} comply with
\begin{equation}
\label{eq:poly_eqn}
\begin{aligned}
G(\lambda)\mfb{x}^* &= 0 \\
(\mfb{x}^*)^TM_\seq(\mfb{x}^*) &= 1 \ \ \forall \seq\in\ses' 
\end{aligned}
\end{equation}
for some unique $\lambda=(\lambda_{\seq})_{\seq\in \ses'}$.
\end{lem}

\begin{pf}
Let $\mathcal{L}(\mfb{x},\lambda) = \mfb{x}^TM_0\mfb{x} + \sum_{\seq\in\ses'}\lambda_\seq(\mfb{x}^TM_\seq\mfb{x} - 1)$ be the Lagrangian for \eqref{eq:subproblem} with Lagrange multipliers $\lambda$. Moreover, $\mathcal{L}(\mfb{x},\lambda)$ can be written as
\begin{equation}
\label{eq:lagrangian}
\mathcal{L}(\mfb{x},\lambda) = \mfb{x}^TG(\lambda)\mfb{x} - \sum_{\seq\in\ses'}\lambda_\seq
\end{equation}
According to \cite[Propositon 3.1.1]{nonlinear_prog}, all regular critical points $\mfb{x}^*$ must satisfy  $(\mfb{x}^*)^TM_\seq\mfb{x}^* = 1, \ \forall \seq\in\ses'$ and
$\nabla_{\mfb{x}} \mathcal{L}(\mfb{x}^*,\lambda) = 0$
or equivalently $G(\lambda)\mfb{x}^* = 0$ for some unique Lagrange multipliers $\lambda$.
\end{pf}

\begin{lem}
\label{le:dual}
Let $\lambda^*$ be any critical point of the dual problem of \eqref{eq:subproblem} and $g(\lambda)$ be its dual objective (in the sense of \cite[Chapter 5]{boyd_convex}). Moreover, let  $(\mfb{x}^*)^TM_0(\mfb{x}^*) - g(\lambda^*)$ be the duality gap between the primal in \eqref{eq:subproblem} and its dual. Then, any regular critical point $\mfb{x}^*$ of the primal has zero duality gap. Consequently, the duality gap is zero if and only if $\lambda^*$ is also a critical point of
\begin{equation}
\label{eq:det}
\begin{aligned}
&\max_{\lambda}\ \ \left(-\sum\nolimits_{\seq\in\ses'}\lambda_\seq\right) \\
&\text{\normalfont s.t. } \text{\normalfont det }G(\lambda)=0 
\end{aligned}
\end{equation}

\end{lem}

\begin{pf}
First, we compute the dual function (see \cite[Page 216]{boyd_convex}) of the objective in \eqref{eq:subproblem} as
$
g(\lambda) = \inf_{\mfb{x}\in\mathbb{R}^n} \mathcal{L}(\mfb{x},\lambda) = \left(\inf_{\mfb{x}\in\mathbb{R}^n} \mfb{x}^TG(\lambda)\mfb{x}\right)-\sum_{\seq\in\ses'}\lambda_\seq
$
where $\mathcal{L}(\mfb{x},\lambda)$ is the Lagrangian given in \eqref{eq:lagrangian}. Therefore, we conclude that
$$
g(\lambda) = \left\{\begin{array}{ll}
     -\sum_{\seq\in\ses'}\lambda_\seq & \text{if } G(\lambda)\succeq 0 \text{ or } \text{det }G(\lambda)=0\\
     -\infty & \text{otherwise}
\end{array}\right.
$$
since the infimum of the form $\mfb{x}^TG(\lambda)\mfb{x}$ is either zero (for $G(\lambda)\succeq 0$ or $\text{det} G(\lambda) = 0$ with $\mfb{x}$ in the kernel of $G(\lambda)$) or $-\infty$ (see \cite[Page 220]{boyd_convex}).
Hence, the dual program results in
\begin{equation}
\label{eq:general}
\begin{aligned}
\max_{\lambda}&\ \  \left(-\sum\nolimits_{\seq\in\ses'}\lambda_\seq\right) \\
\text{\normalfont s.t. }& G(\lambda) \succeq 0
\end{aligned}
\end{equation}
making the constraints explicit (see \cite[Page 224]{boyd_convex}), ignoring the case when $g(\lambda)=-\infty$. Note that due to Lemma \ref{le:xdag} it is true that $G(\lambda^*)\mfb{x}^*=0$ for some $\lambda^*$. Hence, we can multiply \begin{equation}
    \begin{aligned}
    &(\mfb{x}^*)^TG(\lambda^*)\mfb{x}^* = (\mfb{x}^*)^TM_0(\mfb{x}^*) + \sum_{\seq\in\ses'} \lambda_\seq^* (\mfb{x}^*)^TM_\seq(\mfb{x}^*) \\&= (\mfb{x}^*)^TM_0(\mfb{x}^*) + \sum_{\seq\in\ses'}\lambda_\seq^* =  (\mfb{x}^*)^TM_0(\mfb{x}^*) - g(\lambda^*) = 0
    \end{aligned}
\end{equation} 
which results in zero duality gap, only possible if $\lambda^*$ is a critical point of the dual problem by the definition of $g(\lambda)$. Note that zero duality gap plus the condition that $(\mfb{x}^*)^TM_\seq(\mfb{x}^*)=1$ implies $\mfb{x}^*\neq 0$. Then, $G(\lambda^*)\mfb{x}^*=0$ if and only if $\text{det }G(\lambda^*) = 0$ and $\mfb{x}^*$  in the kernel of $G(\lambda^*)$. 
\end{pf}

Using these results, we proceed to show Theorem \ref{th:regular}. Build a Lagrangian for \eqref{eq:det} as 
$
\mathcal{L}'(\lambda,\mu) = -\sum_{\seq\in\ses'}\lambda_\seq + \mu \text{ det }G(\lambda)
$
with Lagrange multiplier $\mu\in\mathbb{R}$. Thus, $\lambda^*$ comply with $\nabla_\lambda \mathcal{L}'(\lambda^*,\mu) = 0$ \cite[Proposition 3.1.1]{nonlinear_prog} and
$
{\partial\mathcal{L}'}/{\partial \lambda_\seq} = -1 + \mu\text{ tr}(G(\lambda^*)^\dagger M_\seq)=0
$
using Jacobi's formula for the derivative of the determinant \cite[Page 29]{horn}.
Thus, $\text{ tr}(G(\lambda^*)^\dagger M_\seq) = 1/\mu$ or equivalently
\begin{equation}
\label{eq:conditions}
\text{ tr}(G(\lambda^*)^\dagger M_\seq) = \text{ tr}(G(\lambda^*)^\dagger M_\seqq), \ \ \forall \seq,\seqq\in\ses'
\end{equation}
This result, in addition to the condition $\text{det }G(\lambda^*)=0$ result in \eqref{eq:poly_eq}. From Lemma \ref{le:dual} we can also conclude that any critical point complies with $(\mfb{x}^*)M_0(\mfb{x}^*)=-\sum_{\seq\in\ses'}\lambda_\seq^*$ since this condition is equivalent to the zero duality gap property of \eqref{eq:det}. Finally, due to Lemma \ref{le:xdag}, we know that $G(\lambda^*)\mfb{x}^*=0$ and this is true only if $\mfb{x}^*$ is in the kernel of $G(\lambda^*)$.

\subsection{Proof of Corollary \ref{cor:nullity}}
\label{app:nullity}

Recall that the rank of $G(\lambda)$ is equivalent to the size of the largest invertible sub-matrix $G(\lambda)$ \cite[Page 12]{horn}. We proceed by contradiction: assume that $\lambda$ doesn't comply with at least one equation \eqref{eq:sub} for some pair $(i,j)$. This would mean that $G_{ij}(\lambda)$ is invertible resulting in the rank of $G(\lambda)$ to be $n-1$. However, the nullity of $G(\lambda)$ was greater than 1. Then, by the rank-nullity theorem \cite[Page 6]{horn} the rank should have been less than $n-1$ leading to a contradiction.

\subsection{Proof of Proposition \ref{prop:non_regular}}
\label{app:non_regular}

Let $V = \{\mf{v}_1,\dots,\mf{v}_r\}$ be arbitrarily $r=n-|\ses'|$ vectors $\mf{v}_1,\dots,\mf{v}_{r}\in\mathbb{R}^n$. Build a matrix $W_V(\mfb{x}) = [W(\mfb{x}),\mf{v}_1,\dots,\mf{v}_{r}]$ . The vectors $\{M_\seq\mfb{x}\}_{\seq\in\ses'}$ are linearly dependent if and only if the matrix $W_V(\mfb{x})$ is singular for any choice of $\mf{v}_1,\dots,\mf{v}_{r}$. Using the Laplace expansion of the determinant of $W_V(\mfb{x})$ \cite[Page 8]{horn} we obtain
$
\text{det } W_V(\mfb{x}) = \sum_{\alpha\in\mathcal{A}} q_\alpha(V)\text{det } W_\alpha(\mfb{x}) 
$
where $q_\alpha(V)$ are polynomials in the components of the vectors in $V$. In order to comply $\text{det } W_V(\mfb{x})=0$ for any set of vectors $V$, then $\text{det } W_\alpha(\mfb{x}) =0$ for all $\alpha\in\mathcal{A}$.

\subsection{Proof of Proposition \ref{th:np}}

\label{app:np}
The proof follows by performing a Karp reduction \cite[Definition 15.15]{np} of the 3-SAT problem, known to be NP-complete \cite[Theorem 15.22]{np}, to an instance of Problem \ref{prob:main}. First, consider an instance of Problem \ref{prob:main} with $A=0,B=I$ and $\Delta^1=\dots=\Delta^D=\Delta, T_f=\alpha \Delta, \alpha\in\mathbb{N}$ with $(I+L^1\Delta),\dots,(I+L^D\Delta)$ being $0-1$ left stochastic matrices. Moreover, let $P_0=0, W_0=0, \Sigma^1=\cdots=\Sigma^D=0$ as well as $Q=0,\lambda_r=0,\lambda_{\mf{x}}=1$ and $Q_f=\diag(c)$ for some $0-1$ vector $c$. Finally, consider the initial condition $\bar{\mf{x}}_0$ to be a $0-1$ vector as well. This results in $\Lambda(\Delta^{p_k}) = I+ L^{p_k}\Delta$ being $0-1$ matrices and as a consequence $\bar{\mf{x}}[\alpha]$ is a $0-1$ vector. Thus, $\mathcal{J}(p) = \bar{\mf{x}}[\alpha]^T\diag(c)\bar{\mf{x}}[\alpha]=c^T\bar{\mf{x}}[\alpha]$. Hence, solving this instance of Problem \ref{prob:main} requires to find a schedule $p_0,\dots,p_{\alpha-1}$ such that the linear function $c^T\mf{x}[\alpha]$ is minimized, or equivalently $-c^T\mf{x}[\alpha]$ is maximized. Thus, the Problem \cite[Problem (P)]{wu} for stochastic matrices, initial condition and vector $c$ all with $0-1$ components can be reduced to Problem \ref{prob:main}. Finally, the proof of \cite[Theorem 3.1]{wu} implies that the 3-SAT problem can be reduced to this instance of \cite[Problem (P)]{wu}.

\subsection{Proof of Proposition \ref{prop:dynprog}}
\label{app:dynprog}

The proof of correctness follows by a direct application of the dynamic programming algorithm \cite[Page 23]{bertsekas2000}. For the complexity, note that the largest amount of recursive calls in Algorithm \ref{algo:dynprog} is obtained when the schedule covering the smallest amount of time $c$ over $[0,T_f]$ is chosen repeatedly. In this case, the depth of recursive calls to \texttt{dynprog} is $\lfloor T_f/c \rfloor$. Hence, due to line 2 in Algorithm \ref{algo:dynprog}, the total number of recursive calls is at most $m^{\lfloor T_f/c\rfloor}$.

\subsection{Proof of Theorem \ref{th:optimal}}
\label{ap:optimal}

For the proof, for any matrix $P$ let $\|P\|$ denote its matrix norm \cite[Section 5.6]{horn} induced by the Euclidean norm. To show Theorem \ref{th:optimal} we provide the following.
\begin{lem}
\label{lem:variation}
Let $p$ be a fixed perception schedule over $[0,T_f]$. Moreover, let any compact sets $B_\mf{\bar{x}}\in\mathbb{R}^n, B_P\in\mathbb{R}^{n\times n}$ and any $\varepsilon>0$. Then, there exists $\delta_{\mf{\bar{x}}},\delta_P>0$ such that for any $\mf{\bar{x}}_0,\mf{\bar{x}}_0'\in B_{\mf{\bar{x}}}$ and any $P,P'\in B_P$ with  $\|\mf{\bar{x}}_0-\mf{\bar{x}}_0'\|<\delta_{\mf{\bar{x}}}$ and $\|P-P'\|<\delta_P$ implies  $|\mathcal{J}(p;\mf{\bar{x}}_0,P_0)-\mathcal{J}(p;\mf{\bar{x}}_0',P_0')|<\varepsilon$.
\end{lem}

\begin{pf}
First, let $r_{\mf{\bar{x}}}=\sup\{\|\bar{\mf{x}}\|:\bar{\mf{x}}\in B_{\bar{\mf{x}}}\}$. Then, we leverage continuity of solutions of \eqref{eq:distro} with respect to initial conditions. Concretely, using the bound in \cite[Page 356 - Equation (9.28)]{khalil} applied to \eqref{eq:distro} for the compact interval $[0,T_f]$ it follows that $\|\mf{\bar{x}}(t)-\mf{\bar{x}}'(t)\|\leq \|\mf{\bar{x}}_0-\mf{\bar{x}}'_0\|\exp(c_{\mf{\bar{x}}}T_f)\leq \delta_{\mf{\bar{x}}}\exp(c_{\mf{\bar{x}}}T_f), \forall t\in[0,T_f]$ for some constant $c_{\mf{\bar{x}}}\in\mathbb{R}$. A similar reasoning leads to obtain $\|P(t)-P'(t)\|\leq \delta_P\exp(c_PT_f), \forall t\in[0,T_f]$ for some constant $c_P\in\mathbb{R}$. Moreover, $|\mf{\bar{x}}(t)^TQ\mf{\bar{x}}(t)-\mf{\bar{x}}(t{)'}^{T}Q\mf{\bar{x}}'(t)| = |(\mf{\bar{x}}(t)+\mf{\bar{x}}'(t))^TQ(\mf{\bar{x}}(t)-\mf{\bar{x}}'(t))|\leq 2r_{\mf{\bar{x}}}\|Q\|\delta_{\mf{\bar{x}}}\exp(c_{\bar{\mf{x}}}T_f)$ using the matrix norm properties in \cite[Section 5.6]{horn}. Similarly, $|\tr(QP(t))-\tr(QP'(t))|=|\tr(Q(P(t)-P'(t)))|\leq n\|Q(P(t)-P'(t))\|\leq n\|Q\|\delta_P\exp(c_PT_f)$. Hence, using these results on $e:=|\mathcal{J}(p;\mf{\bar{x}}_0,P_0)-\mathcal{J}(p;\mf{\bar{x}}'_0,P'_0)|$ which using \eqref{eq:cost_functional_det} leads to conclude that for appropriate (sufficiently small) $\delta_{\mf{\bar{x}}},\delta_P>0$:
$
e\leq 2\lambda_{\mf{x}}r_{\mf{\bar{x}}}\delta_{\mf{\bar{x}}}(\|Q\|+\|Q_f\|)\exp(c_{\mf{\bar{x}}}T_f)+ n\lambda_{\mf{x}}\delta_P(\|Q\|+\|Q_f\|)\exp(c_PT_f))\leq \varepsilon$
\end{pf}

We proceed to show Theorem \ref{th:optimal}. First, given $\varepsilon>0$ and $B_{\mf{\bar{x}}},B_p$ choose $\delta_{\mf{\bar{x}}},\delta_P$ as in Lemma \ref{lem:variation} and note that this values are independent on initial conditions as long as they lie in $B_{\mf{\bar{x}}},B_P$. Hence, there is $N>0$ such that we can partition the compact set $B_{\mf{\bar{x}}}$ into regions $B_{\mf{\bar{x}}}^1,\dots,B_{\mf{\bar{x}}}^N$ which cover all $B_{\mf{\bar{x}}}$ and $\|\mf{\bar{x}}-\mf{\bar{x}}'\|\leq \delta_{\mf{\bar{x}}}, \forall {\mf{\bar{x}}},{\mf{\bar{x}}}'\in B_{\mf{\bar{x}}}^i, \forall i\in\{1,\dots,N\}$. A similar partition of $N$ sections can be performed for $B_P$ as well. Now, choose any $\mf{\bar{x}}^i_0\in B_{\mf{\bar{x}}}^i,P_0^i\in B_P^i$ for partition $i$ and denote with $p^{*i}$ the optimal schedule in this case for the general setting of Problem \ref{prob:main}. As a consequence of Lemma \ref{lem:variation}, the cost of $p^{*i}$ for any $\mf{\bar{x}}_0'\in B_\mf{\bar{x}}^i, P_0'\in B_P^i$ differs from the optimal one in that partition by at most $\varepsilon$. Now, let $\Gamma$ be any admissible set of schedules. Moreover, since $p^{*i}$ is a solution of Problem \ref{prob:main}, then it must be a stabilizing perception schedule for $\mf{\bar{x}}[0]=\mf{\bar{x}}_0^i$.
Thus, there exists $\ell^i<\infty$ (sufficiently large) with $T_f\leq \sum_{k=0}^{\ell^i-1}\Delta^{p_k^{*i}}$ such that $\|\mf{\bar{x}}[\ell^i]\|^2<\|M_0\|^{-1} \min_{\gamma\in\Gamma}(\mfb{x}_0^i)^TM_\seq\mfb{x}_0^i$ for the truncated schedule $\gamma^i=\{p^{*i}_k\}_{k=0}^{{\ell^i}-1}$. This last relation implies $(\mf{\bar{x}}_0^i)^TM_{\gamma^i}\mf{\bar{x}}_0^i=(\Lambda^{\gamma^i}\mf{\bar{x}}_0^i)^TM_0(\Lambda^{\gamma^i}\mf{\bar{x}}_0^i) = \mf{\bar{x}}[\ell^i]^TM_0\mf{\bar{x}}[\ell^i]\leq\|M_0\|\|\mf{\bar{x}}[\ell^i]\|^2<  \min_{\gamma\in\Gamma}(\mfb{x}_0^i)^TM_\seq\mfb{x}_0^i$. Hence, $\Gamma^i=\Gamma\cup\{\gamma^i\}$ is admissible and complies $\gamma^\sstar(\mf{\bar{x}}_0^i
;\Gamma^i)=\gamma^i$ by \eqref{eq:sw_law} and similarly for any $\mf{\bar{x}}_0\in B_{\mf{\bar{x}}}^i$ and $\delta_{\bar{\mf{x}}}$ sufficiently small.
Therefore, the result follows by choosing $m=N$ and all $\Gamma^1,\dots,\Gamma^m$ constructed in this way, one for each partition. The reason is that for any $\mf{\bar{x}}_0\in B_\mf{\bar{x}}, P_0\in B_P$, a schedule with cost differing from the optimal one by at most $\varepsilon$ is always evaluated in Algorithm \ref{algo:dynprog}.

\bibliographystyle{plain}

\end{document}